\documentclass[floatfix,prd,epsfig,nofootinbib,superscriptaddress,onecolumn,amssymb]{revtex4}

\usepackage{slashed}
\usepackage{slashed}
\usepackage{graphicx,color}
\usepackage{epsfig}
\usepackage{subfigure}
\usepackage{epsfig}
\usepackage{dcolumn}
\usepackage{bm}
\usepackage{color}
\usepackage{hyperref}
\usepackage{mathrsfs}
\usepackage{lmodern,textcomp}

\def\lsim{\mathrel{\rlap{\lower4pt\hbox{\hskip1pt$\sim$}}
    \raise1pt\hbox{$<$}}}         
\def\gsim{\mathrel{\rlap{\lower4pt\hbox{\hskip1pt$\sim$}}
    \raise1pt\hbox{$>$}}}         
    
    \newcommand{\nc}{\newcommand}  

\nc{\ra}{\rightarrow}  
\nc{\slsh}{\slash\hspace*{-0.22cm}}

\def\to{\rightarrow}
\def\Re{{\cal R \mskip-4mu \lower.1ex \hbox{\it e}\,}}
\def\Im{{\cal I \mskip-5mu \lower.1ex \hbox{\it m}\,}}
\def\be{\begin{equation}}
\def\ee{\end{equation}}
\def\bea{\begin{eqnarray}}
\def\eea{\end{eqnarray}}
\def\bit{\begin{itemize}}
\def\eit{\end{itemize}}
\nc{\eref}[1]{(\ref{#1})}
\nc{\Eref}[1]{Eq.~(\ref{#1})}

\nc{\vev}[1]{ \left\langle {#1} \right\rangle }
\nc{\bra}[1]{ \langle {#1} | }
\nc{\ket}[1]{ | {#1} \rangle }
\nc{\fb}{\,{\rm fb}^{-1}}
\nc{\ev}{{\rm eV}}
\nc{\kev}{{\rm keV}}
\nc{\Mev}{{\rm MeV}}
\nc{\gev}{{\rm GeV}}
\nc{\tev}{{\rm TeV}}
\nc{\mev}{{\rm MeV}}

\def\ee{e^+e^-}

\def\msb{{\bar{\ssstyle M \kern -1pt S}}}

\def\eps{\epsilon}

\newcommand{\NIST}{National Institute of Standards and Technology (NIST) \renewcommand{\NIST}{NIST}}
\newcommand{\LHC}{Large Hadron Collider (LHC) \renewcommand{\LHC}{LHC}}
\newcommand{\LRP}{Long Range Plan (LRP) \renewcommand{\LRP}{LRP}}
\newcommand{\BSM}{Beyond-Standard-Model (BSM) \renewcommand{\BSM}{BSM}}
\newcommand{\CP}{charge conjugation and parity (CP) \renewcommand{\CP}{CP}}
\newcommand{\FNPB}{undamental Neutron Physics Beamline) (FNPB) \renewcommand{\FNPB}{FNPB}}
\newcommand{\EDM}{Electric Dipole Moment (EDM) \renewcommand{\EDM}{EDM}}
\newcommand{\LANL}{Los Alamos National Laboratory (LANL) \renewcommand{\LANL}{LANL}}
\newcommand{\SM}{Standard Model (SM) \renewcommand{\SM}{SM}}
\newcommand{\CKM}{Cabibbo-Kobayashi-Maskawa (CKM) \renewcommand{\CKM}{CKM}}

\begin{document}

\title{Fundamental Neutron Physics: a White Paper on Progress and Prospects in the US}

\author{R.~Alarcon}
\affiliation{Arizona State University}

\author{A.~Aleksandrova}
\affiliation{California Institute of Technology}

\author{S.~Bae{\ss}ler}\email{baessler@virginia.edu}
\affiliation{University of Virginia}
\affiliation{Oak Ridge National Laboratory}

\author{D.~H.~Beck}
\affiliation{University of Illinois, Urbana-Champagne}

\author{T.~Bhattacharya}
\affiliation{Los Alamos National Laboratory}

\author{M.~Blatnik}
\affiliation{California Institute of Technology}

\author{T.~J.~Bowles}
\affiliation{Los Alamos National Laboratory}
\affiliation{University of Washington}

\author{J.~D.~Bowman}
\affiliation{Oak Ridge National Laboratory}

\author{J.~Brewington}
\affiliation{University of Kentucky}

\author{L.~J.~Broussard}\email{brousssardlj@ornl.gov}
\affiliation{Oak Ridge National Laboratory}

\author{A.~Bryant}
\affiliation{University of Virginia}

\author{J.~F.~Burdine}
\affiliation{University of Illinois, Urbana-Champagne}

\author{J.~Caylor}
\affiliation{Syracuse University}

\author{Y.~Chen}
\affiliation{Indiana University}

\author{J.~H.~Choi}
\affiliation{North Carolina University}

\author{L.~Christie}
\affiliation{University of Tennessee, Knoxville}

\author{T.~E.~Chupp}\email{chupp@umich.edu}
\affiliation{University of Michigan}

\author{V.~Cianciolo}
\affiliation{Oak Ridge National Laboratory}

\author{V.~Cirigliano}
\affiliation{University of Washington}

\author{S.~M.~Clayton}
\affiliation{Los Alamos National Laboratory}

\author{B.~Collett}
\affiliation{Hamilton College}

\author{C.~Crawford}
\affiliation{University of Kentucky}

\author{W.~Dekens}
\affiliation{University of Washington}

\author{M.~Demarteau}
\affiliation{Oak Ridge National Laboratory}

\author{D.~DeMille}
\affiliation{University of Chicago}
\affiliation{Argonne National Lab}

\author{G.~Dodson}
\affiliation{Massachusetts Institute of Technology}

\author{B.~W.~Filippone}\email{bradf@caltech.edu}
\affiliation{California Institute of Technology}
\affiliation{Oak Ridge National Laboratory}

\author{N.~Floyd}
\affiliation{University of Kentucky}

\author{N.~Fomin}
\affiliation{University of Tennessee, Knoxville}

\author{J~Fry}
\affiliation{Eastern Kentucky University}

\author{K.~Fuyuto}
\affiliation{Los Alamos National Laboratory}

\author{S.~Gardner}
\affiliation{University of Kentucky}

\author{R.~Godri}
\affiliation{University of Tennessee, Knoxville}

\author{R.~Golub}
\affiliation{North Carolina State University}

\author{F.~Gonzalez}
\affiliation{Oak Ridge National Laboratory}

\author{G.~L.~Greene}
\affiliation{University of Tennessee, Knoxville}
\affiliation{Tulane University}

\author{V.~Gudkov}
\affiliation{University of South Carolina}

\author{R.~Gupta}
\affiliation{Los Alamos National Laboratory}

\author{J.~Hamblen}
\affiliation{University of Tennessee-Chattanooga}

\author{L.~Hayen}\email{lmhayen@nscu.edu}
\affiliation{North Carolina State University}

\author{C.~Hendrus}
\affiliation{University of Michigan}
\affiliation{Ohio State University}

\author{K. Hickerson}
\affiliation{California Institute of Technology}

\author{F.~B.~Hills}
\affiliation{University of Michigan}

\author{A.~T.~Holley}
\affiliation{Tenessee Technological University}

\author{S.~Hoogerheide}
\affiliation{National Institute of Standards and Technology}

\author{M.~Hubert}
\affiliation{National Institute of Standards and Technology}

\author{P.~R.~Huffman}
\affiliation{North Carolina State University}

\author{S.~K.~Imam}
\affiliation{University of Tennessee, Knoxville}

\author{T.~M.~Ito}\email{ito@lanl.gov}
\affiliation{Los Alamos National Laboratory}

\author{L.~Jin}
\affiliation{University of Connecticut}

\author{G.~Jones}
\affiliation{Hamilton College}

\author{A.~Komives}
\affiliation{DePauw University}

\author{E.~Korobkina}
\affiliation{North Carolina State University}

\author{W.~Korsch}
\affiliation{University of Kentucky}

\author{K.~K.~H.~Leung}
\affiliation{Montclair State University}

\author{C.-Y.~Liu}\email{chenyliu@illinois.edu}
\affiliation{University of Illinois, Urbana-Champagne}

\author{K.-F. Liu}
\affiliation{University of Kentucky}

\author{J.~C.~Long}
\affiliation{University of Illinois, Urbana-Champagne}

\author{D.~Mathews}
\affiliation{Oak Ridge National Laboratory}

\author{A.~Mendelsohn}
\affiliation{University of Manitoba}

\author{E.~Mereghetti}\email{emereghetti@lanl.gov}
\affiliation{Los Alamos National Laboratory}

\author{P.~Mohanmurthy}
\affiliation{Massachusetts Institute of Technology}

\author{C.~L.~Morris}
\affiliation{Los Alamos National Laboratory}

\author{P.~Mueller}
\affiliation{Oak Ridge National Laboratory}

\author{H.~P.~Mumm}\email{hans.mumm@nist.gov}
\affiliation{National Institute of Standards and Technology}

\author{A.~Nelsen}
\affiliation{University of Kentucky}

\author{A.~Nicholson}
\affiliation{University of North Carolina}

\author{J. Nico}
\affiliation{National Institute of Standards and Technology}

\author{C.~M.~O'Shaughnessy}
\affiliation{Los Alamos National Laboratory}

\author{P.~A.~Palamure }
\affiliation{University of Kentucky}

\author{S.~Pastore}
\affiliation{Washington University in St. Louis}

\author{R.~W.~Pattie~Jr.}
\affiliation{East Tennessee State University}

\author{N.~S.~Phan}
\affiliation{Los Alamos National Laboratory}

\author{J.~A.~Pioquinto}
\affiliation{University of Virginia}

\author{B.~Plaster}
\affiliation{University of Kentucky}
\affiliation{Oak Ridge National Laboratory}

\author{D.~Po\v{c}ani\'c}
\affiliation{University of Virginia}

\author{H.~Rahangdale}
\affiliation{University of Tennessee, Knoxville}

\author{R.~Redwine}
\affiliation{Massachusetts Institute of Technology}

\author{A.~Reid}
\affiliation{Tennessee Technological University}

\author{D.~J.~Salvat}
\affiliation{Indiana University}

\author{A.~Saunders}
\affiliation{Oak Ridge National Laboratory}

\author{D.~Schaper}
\affiliation{Los Alamos National Laboratory}

\author{C.-Y. Seng}\email{seng@frib.msu.edu}
\affiliation{Michigan State University}
\affiliation{University of Washington}

\author{M.~Singh}
\affiliation{Los Alamos National Laboratory}

\author{A.~Shindler}
\affiliation{Michigan State University}

\author{W.~M.~Snow}\email{wsnow@indiana.edu}
\affiliation{Indiana University}

\author{Z.~Tang}
\affiliation{Los Alamos National Laboratory}

\author{A.~Walker-Loud}
\affiliation{Lawrence Berkeley National Laboratory}

\author{D.~K.-T.~Wong}
\affiliation{Indiana University}

\author{F. Wietfeldt}
\affiliation{Tulane University}

\author{A.~R.~Young}\email{aryoung@nscu.edu}
\affiliation{North Carolina State University}
\maketitle

\newpage
\section*{Executive Summary}
\vskip -.1in

Fundamental neutron physics, combining precision measurements and theory, probes particle physics at short range with reach well beyond the highest energies probed by the LHC. Significant US efforts are underway that will probe BSM CP violation with orders of magnitude more sensitivity, provide new data on the Cabibbo anomaly, more precisely measure the neutron lifetime and decay, and explore hadronic parity violation. World-leading results from the US Fundamental Neutron Physics community since the last Long Range Plan, include the world's most precise measurement of the neutron lifetime from UCN$\tau$, the final results on the beta-asymmetry from UCNA and  new results on hadronic parity violation from the NPDGamma and n-${}^3$He runs at the FNPB (Fundamental Neutron Physics Beamline), precision measurement of the radiative neutron decay mode and n-${}^4$He at NIST. US leadership and discovery potential are ensured by the development of new high-impact experiments including BL3, Nab, LANL nEDM and nEDM@SNS. 
On the theory side, the last few years have seen results for the neutron EDM from the QCD $\theta$ term, a factor of two reduction in the uncertainty for inner radiative corrections in beta-decay which impacts CKM unitarity, and progress on {\it ab initio} calculations of nuclear structure for medium-mass and heavy nuclei which can eventually improve the connection between nuclear and nucleon EDMs. 

\vskip .1in
In order to maintain this exciting program and capitalize on past investments while also pursuing new ideas and building US leadership in new areas, the Fundamental Neutron Physics community has identified a number of priorities and opportunities for our sub-field covering the time-frame of the last Long Range Plan (LRP) under development. These are:

\begin{itemize}
  \item Funding for completion of the construction of the nEDM@SNS apparatus and the start of data-taking
  \item Investment in additional funding to support research and beamline operations, including additional personnel, for FNPB, the NIST fundamental neutron physics beamlines, and the LANL UCN source to maintain leadership in the field, ensuring adequate resources to run experiments, improve capabilities and provide continuity
  \item Increased support for theoretical groups that are involved in all aspects of fundamental neutron physics research, from phenomenology, to effective field theories, hadronic physics and lattice QCD, expanding connections with the high-energy physics and nuclear structure communities
\item Development of new funding mechanisms to support R\&D activities focused on future experiments and capabilities, such as a next-generation UCN source and high flux, high polarization and high uniformity neutron beam polarizers for cold neutron beams. In this regard we note that the Fundamental Neutron Physics community does not have a mission-centered neutron facility and thus faces significant challenges in carrying out this research
\end{itemize}











\clearpage
\section{Introduction}

The Standard Model (SM) of Elementary Particles is an extremely successful theory, which has passed a large number of stringent experimental tests,  
both at high- and low-energy. 
With the discovery of the Higgs boson at the Large Hadron Collider \cite{ATLAS:2012yve,CMS:2012qbp}, the SM is now complete. 
We know, on the other hand, that the SM is not the definitive theory of Nature, as it does not accommodate 
neutrino masses, it cannot successfully generate the observed matter-antimatter asymmetry in the Universe, and does not have a viable dark matter candidate. 
In addition to these longstanding open questions,
the last few years have witnessed the emergence of significant tensions in tests of the unitarity of the Cabibbo-Kobayashi-Maskawa (CKM) matrix
\cite{Seng:2018yzq,Czarnecki:2019mwq,Shiells:2020fqp,Hardy:2020qwl}, tests
of lepton universality in semileptonic $B \rightarrow K \ell^+ \ell^-$ decays \cite{LHCb:2021trn},
and measurements of the anomalous magnetic moment of the muon
 \cite{Muong-2:2006rrc,Muong-2:2021ojo}, just to name a few. These anomalies could signal the first cracks in the SM edifice. 
To address the shortcomings of the SM,
a vigorous high-energy experimental program is in place, involving experiments at the Large Hadron Collider (LHC), at machines dedicated to the exploration of the flavor sector of the SM 
\cite{Belle-II:2018jsg}, and  the next generation of neutrino oscillation experiments \cite{DUNE}.
Alternatively, precision low-energy experiments with neutrons are uniquely positioned to  answer some of the most pressing open questions in the SM and in some cases have significantly higher mass-scale reach than possible at the LHC. Such experiments thus provide highly competitive and complementary information in the search for Beyond-Standard-Model (BSM) physics. 
\\

{\bf Neutrons and the matter-antimatter asymmetry in the Universe. } 
Neutrons are sensitive probes
of two approximate symmetries of the SM, 
baryon number (B) and 
charge conjugation and parity (CP), 
whose violation is necessary to satisfy 
two of the three Sakharov conditions for the dynamical generation of a baryon asymmetry in the Universe (BAU)\footnote{In the SM, $B$ is broken by an anomaly. At the electroweak scale, $B$-violating processes mediated by sphalerons are not negligible   \cite{Kuzmin:1985mm}, and play an important role in electroweak baryogenesis scenarios.  } \cite{Sakharov:1967dj}.
Being neutral, long lived and spin 1/2, the neutron is the simplest hadronic system that can be used to search for a permanent electric dipole moment (EDM) \cite{Purcell:1950zz,Smith:1957ht}, 
a signal  of
time-reversal-violation \cite{Landau1957} (and thus of CP-violation) largely insensitive to the CP-violation (CPV) induced by the phase of the CKM matrix. 
The next generation of experiments will improve existing bounds by one to two orders of magnitude, pushing the sensitivity to new sources of CPV well into the multi-TeV scale. 
Neutron transmission experiments 
can also provide powerful constraints on the time-reversal-violating nucleon-nucleon potential  
\cite{Gudkov:1990tb,Bowman:2014fca},
competitive with and complementary to the neutron, atomic and molecular EDMs \cite{Bowman:2014fca}.
An observation of the neutron EDM or
 a positive signal in a $\beta$ decay experiment or at NOPTREX
could reveal the new CPV sources that are needed for baryogenesis 
but are difficult to probe directly in high energy experiments  \cite{Cirigliano:2016nyn,Alioli:2017ces,Cirigliano:2019vfc,Gritsan:2022php}.
In addition, searches for T-violating correlations in neutron $\beta$ decay continue to have the potential to explore model space not constrained by current EDM searches~\cite{RAMSEYMUSOLF2021136136,PhysRevD.87.116012}. Thus neutron experiments, in conjunction with LHC searches for new degrees of freedom that are able to trigger a first order phase transition 
\cite{Morrissey:2012db,Huang:2017jws, Ramsey-Musolf:2019lsf,Fuyuto:2019svr,Bell:2019mbn,Wang:2022dkz},
can address the fundamental question of whether the matter-antimatter asymmetry is generated close to the electroweak scale (electroweak baryogenesis).

Neutron experiments can also address the baryon number violation that is required in the Sakharov conditions. As a neutral particle, the neutron could have a small Majorana mass term \cite{Kuzmin:1970nx}, which violates B by two units and cause the oscillation of neutrons into antineutrons. 
The current limit on free neutron oscillation time $\tau_{n \bar n} \gtrsim 10^8$ s can be converted into new physics scales of $10^2$-$10^3$ TeV, and upcoming experiments at the European Spallation Source will probe parameter space relevant to low-scale baryogenesis scenarios in which the BAU is induced 
by the B violating decays of new particles that mediate $n$-$\bar n$
oscillations
\cite{Babu:2006xc,Grojean:2018fus}.
\\

{\bf Neutrons as probes of BSM physics at the TeV scale.}
Nuclear $\beta$ decays have been instrumental in the construction of the SM. With experimental precision approaching the permille level and robust theoretical predictions, $\beta$ decays continue to be highly competitive with the constraints from high-energy colliders \cite{Cirigliano:2012ab,Cirigliano:2013xha,Gonzalez-Alonso:2018omy}.
Furthermore, a recent reevaluation of
the ``inner radiative correction'' \cite{Seng:2018yzq, Czarnecki:2019mwq,Shiells:2020fqp,Hardy:2020qwl}, and progress on the lattice input for the extraction of $V_{us}$
and $V_{us}/V_{ud}$ from kaon decays \cite{FlavourLatticeAveragingGroupFLAG:2021npn} have led to a $\sim 3\sigma$ tension in the unitarity of the first row of the CKM matrix, which could be explained by new left- or right-handed couplings of the $W$ boson to quarks and leptons
arising at scales of $\gtrsim 10$ TeV. Such 
couplings can evade constraints from electroweak precision data and are hard to probe directly at the LHC \cite{Cirigliano:2021yto,Cirigliano:2022yyo},
providing an outstanding example of the need of complementary approaches to BSM searches.
The neutron is an ideal system for high precision $\beta$ decays, as the theoretical interpretation is not affected by nuclear theory uncertainties.
Adopting the single best measurement of the neutron lifetime from UCN$\tau$~\cite{UCNt:2021pcg} and 
of the ratio $\lambda = g_A/g_V$ from PERKEO III~\cite{Markisch:2018ndu} the total uncertainty on the extraction of $V_{ud}$ from neutron decay  is already comparable to superallowed $\beta$ decays. 
 Using the Particle Data Group global averages as the standard, matching the $0^+ \rightarrow 0^+$ accuracy will require a factor of two improvement in the uncertainty for the lifetime, 
 $\Delta\tau_n \sim 0.3$~s, 
 and a factor of 3 for $\lambda$, $\Delta\lambda/|\lambda|\sim 0.03\,\%$. 
These goals can be achieved by the US experimental program.
In addition to unitarity tests, the comparison of the experimental value of $\lambda$ with high-precision lattice QCD calculations \cite{Chang:2018uxx,Gupta:2018qil,Aoki2021} provides a uniquely sensitive probe of right-handed charged-currents \cite{Cirigliano:2022hob,Hayen:2020cxh}.
Sub-permille measurements of the Fierz interference term, $b$, can probe  scalar and tensor interactions at a level competitive with the LHC and meson decays.

In addition to reach far beyond the electroweak scale, neutrons are key to a better understanding of the SM.
The nucleon-nucleon weak interaction is one of the most poorly-understood sectors of the electroweak theory.
Low-energy hadronic parity violating experiments provide the opportunity to test our ability to trace symmetry-violating effects of the known quark-quark electroweak interaction 
from the electroweak scale,
across nonperturbative strong interaction distance scales at and above $\Lambda_{QCD}$, all the way down to nuclear, atomic, and molecular scales. 
Since a similar exercise must also be performed for other BSM searches for symmetry-violating low energy nuclear observables such as electric dipole moments and neutrinoless double beta decay, hadronic parity violating experiments are an ideal testing ground of the theoretical tools at our disposal. Ongoing work towards lattice calculations~\cite{Kurth:2015cvl} of nucleon-nucleon (NN) weak amplitudes inhabits the ``computational frontier" of the SM~\cite{Drischler:2019xuo}.
\\

{\bf Neutrons and new weakly-interacting particles.}
Another possibility, which is gaining more and more attention \cite{Battaglieri:2017aum}, is that physics beyond the SM is light and very weakly coupled.
The extended symmetries present in many theories beyond the SM (including string theories) are typically broken at some high energy scale, leading to new weakly-coupled light particles with relatively long-range interactions~\cite{Weinberg72, Arvanitaki2010, Jae10}. Spin $0$ and spin $1$ boson exchange generates several (in general spin-dependent) interactions in the nonrelativistic limit~\cite{Dob06, Fadeev2019b}. Effective field theory treatments of dark matter ``quantum dark forces”~\cite{Fichet2017, Brax2018} can also be parametrized in a similar way. Slow neutron interactions have been exploited in several searches for possible new weakly coupled interactions of various types, including chameleon and symmetron dark energy fields, light $Z^{'}$ bosons, in-matter gravitational torsion and nonmetricity of spacetime, axion-like particles, short-range modifications of gravity, and exotic parity-odd interactions~\cite{Sponar2021}, complementing similar experiments performed with atoms and molecules~\cite{Safronova2018}. These neutron experiments provide useful constraints on a host of exotic BSM interactions and can be greatly improved in sensitivity.


\section{Progress and Prospects}
\subsection{CP violation}

\noindent
\subsubsection{Motivation}

The origin of the matter-antimatter asymmetry in the Universe is one of the most pressing open problems in fundamental physics. 
The SM  is missing two key ingredients to generate a baryon asymmetry compatible with observation: 
with a Higgs mass of 125 GeV,
the SM electroweak (EW) phase transition does not provide enough departure from thermal equilibrium and the CP violation induced by the phase of the CKM matrix is too small \cite{Gavela:1993ts,Gavela:1994ds,Gavela:1994dt,Huet:1994jb}. 
Neutron experiments offer a unique window onto 
new sources of CP-violation, as they combine very high sensitivity 
with essentially no background from CPV sources in the SM.
Searches for an electric dipole moment of the neutron are by now a classic test of T-violation \cite{Purcell:1950zz,Smith:1957ht}. The current bound on the neutron EDM (nEDM),
$| d_n | < 1.8\times 10^{-26}$~$e\cdot$cm \cite{Abel2020}, 
can be naively converted into sensitivity to new physics scales in the $1$-$100$ TeV range, depending on whether EDMs are induced at the tree or loop level, and generically rules out
new physics at the TeV scale, with couplings to light SM fermions and $\mathcal O(1)$ CPV phases \cite{Pospelov:2005pr,Morrissey:2012db}.
Even if CPV arises predominantly via couplings to heavy SM degrees of freedom, such as the top quark, the Higgs or weak gauge bosons,
EDMs provide constraints on the CP phases that are typically much stronger than direct collider probes
\cite{Brod:2013cka,Chien:2015xha,Cirigliano:2016nyn,Cirigliano:2019vfc,Brod:2022bww,Gritsan:2022php}. The next generation of experiments will improve the nEDM bounds by two orders of magnitude
and will probe regions in parameter space of interest to several electroweak baryogenesis scenarios 
\cite{Li:2008ez,Cirigliano:2009yd,Morrissey:2012db}.

While the observation
of an nEDM will be  paradigm-shifting, 
by itself it will not be sufficient to discriminate among various BSM models and thereby among different baryogenesis scenarios. 
To achieve this goal, it is
first of all necessary to observe 
CPV in complementary systems, so that  
the main features of CPV at low energy can be identified. Searches for EDMs 
of leptons, proton, light ions, atoms and molecules are generally orthogonal to the nEDM, as they probe different sets of low-energy CPV couplings
\cite{Engel:2013lsa,Alarcon:2022ero}. 
Searches of  T-odd correlations
in the transmission of polarized neutrons through polarized targets are sensitive to the nucleon-nucleon P-odd T-odd potential 
\cite{Bowman:2014fca}, and are thus also orthogonal to the nEDM.
Exploiting enhancements due to small energy splitting between states of opposite parity in heavy nuclei, the NOPTREX experiment can probe T-odd pion-nucleon couplings competitively with the neutron and atomic EDMs
\cite{Bowman:2014fca}.
In addition, measurements or bounds on the T-odd $D$ coefficient in neutron $\beta$ decay have the potential to explore the parameter space of well motivated models, such as the minimal Left-Right Symmetric Model (mLRSM), not excluded by  
current EDM searches~\cite{RAMSEYMUSOLF2021136136,PhysRevD.87.116012}.

Secondly, to understand the implications of low-energy searches
of CPV, it is necessary to
systematically connect  nucleon- and nuclear-level CPV with microscopic theories at the quark-gluon level.
Because interactions of quarks and gluons are strong at these low energies,
this is a challenging task which requires nonperturbative techniques,
and constitutes a major source of uncertainty in quantitatively comparing EDM and collider sensitivity to BSM physics. Over the last decade, Lattice QCD calculations have gradually replaced QCD models
in the evaluation of the nucleon-level couplings induced by quark-gluon CPV operators \cite{Abramczyk:2017oxr,Bhattacharya:2021lol,Bhattacharya:2022whc,Dragos:2019oxn,Alexandrou:2020mds}, but significant challenges remain before the achievement of  
results with control on all sources of systematic errors.
Similarly,
significant effort from the nuclear theory community will be required to achieve  \textit{ab initio} calculations of atomic and molecular EDMs with controlled uncertainties  \cite{Engel:2013lsa,Dobaczewski:2018nim,Yanase:2020agg,Alarcon:2022ero}.

\begin{table}[t]
	\begin{centering}
		\includegraphics[width=1.0\linewidth]{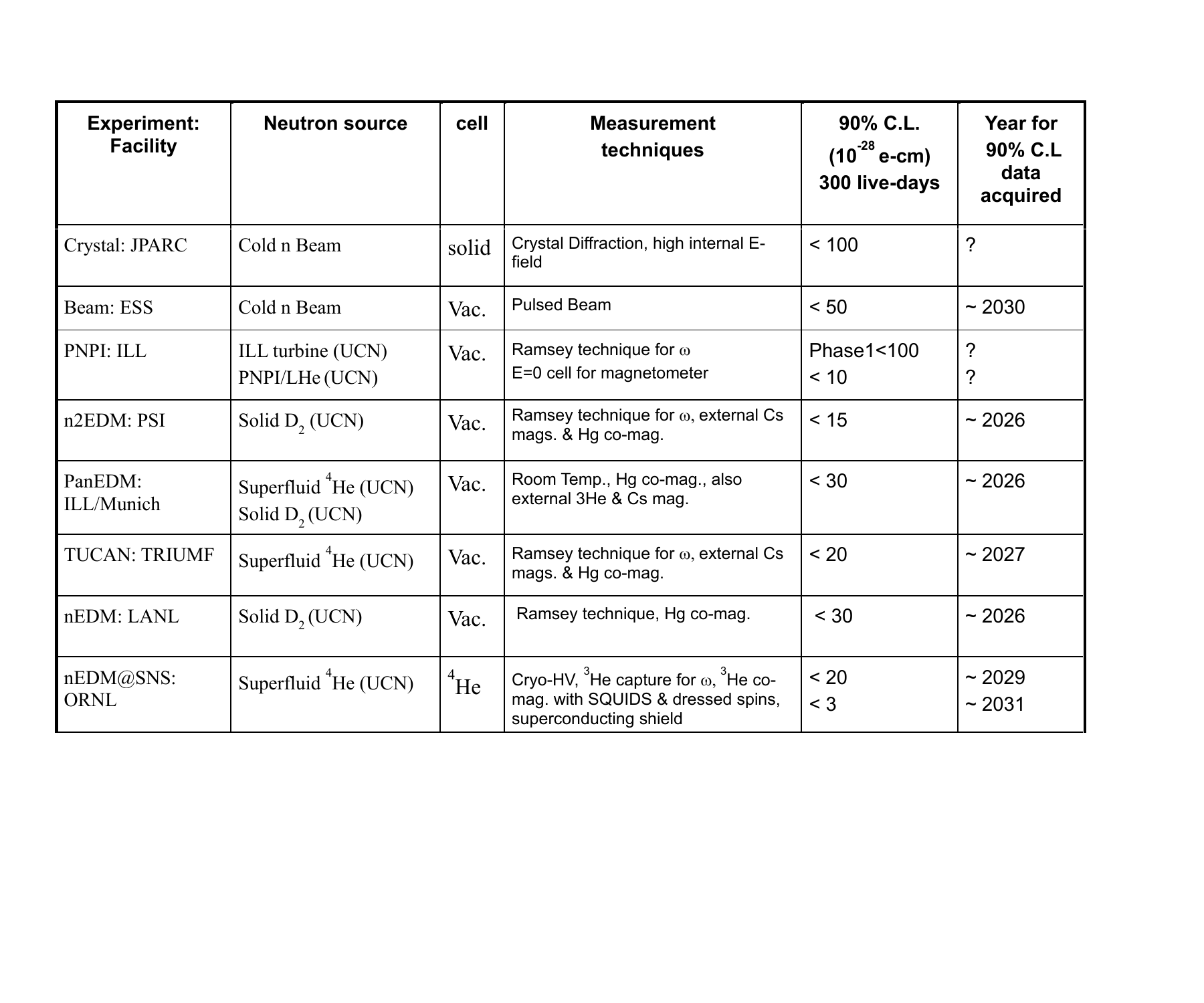}
		\par\end{centering}
	\vskip -1.5in\caption{Worldwide efforts on new neutron EDM experiments. The present 90\,\% confidence level upper limit on the nEDM is 1.8$\times 10^{-26}$ e-cm
\label{fig:nedm_world}
	}
\end{table}

\subsubsection{International context}
\noindent{\bf nEDM}

The present nEDM limit, $| d_n | < 1.8\times 10^{-26}$~$e\cdot$cm (90\,\% C.L.), was set by an experiment performed at the Paul Scherrer Institut (PSI) in Switzerland~\cite{Abel2020}. This experiment inherited the apparatus previously used in the Sussex-ILL experiment~\cite{Baker:PhysRevLett.97.131801,Pendlebury:2015} performed at the Institut Laue-Langevin (ILL) in France, made significant improvements to it, and used it at the PSI UCN source. This result represents a moderate statistical improvement and a factor of 5 reduction in systematic uncertainty compared to the previous limit~\cite{Baker:PhysRevLett.97.131801,Pendlebury:2015}. 

Currently, in addition to the two US efforts (nEDM@SNS and LANL nEDM experiments) that are discussed below, there are two efforts outside the US using cold neutron beams and at least four efforts developing new nEDM experiments using UCN. An overview of all of these experiments is shown in Table \ref{fig:nedm_world}. The cold beam experiments are new concepts that are in the R\&D stage. The non-US UCN experiments include: i)~the PNPI experiment at ILL, ii)~the n2EDM experiment at PSI, iii)~the PanEDM experiment at ILL, and iv)~the TUCAN experiment at TRIUMF in Canada. All of these experiments, along with the LANL nEDM experiment mentioned below, are based on Ramsey's separated oscillatory field method with two precession chambers at room temperature and aim to improve the sensitivity by approximately an order-of-magnitude. The nEDM@SNS experiment, based on a concept first discussed by Golub and Lamoreaux (Ref.~\cite{Golub_Lamoreaux_1994}), uses two different approaches to extract the E-field-dependent frequency shift and aims to improve the sensitivity by approximately two orders-of-magnitude.

\subsubsection{Progress and prospects}
\vskip .2in
\noindent{\bf Theory}

The observation of 
an nEDM in the next generation of experiments will be a clear sign of physics beyond the SM. To understand the implications of such an observation, or of a tighter nEDM bound, on the underlying physics, and to identify the fundamental mechanism of CP-violation, it is  necessary to solve an ``inverse problem'', that is to connect a hadronic observable, the nEDM, to the dynamics of quarks and gluons, to physics at the electroweak scale and models of new physics which might involve scales much larger than the electroweak.
To solve the inverse problem, it is important to explore the correlations between the nEDM and other low-energy CPV observables and with experiments at high-energy colliders, from the Large Hadron Collider to the next generation of $B$ factories.

Since the last Long Range Plan, 
a significant amount of work has been carried out to
systematically connect effective field theory descriptions of BSM physics valid at the EW scale, such as the SM Effective Field Theory (SMEFT) \cite{Buchmuller:1985jz, Grzadkowski:2010es}, with the phenomenology of EDMs, in order to provide a clear picture of the possible blind directions not constrained by EDMs, and thus of the complementary probes of flavor-diagonal CPV to be investigated at the LHC (for some discussion, we refer to the Snowmass white paper  \cite{Gritsan:2022php}, and references therein). 
The framework can then be used to constrain explicit models, such as the mLRSM \cite{Bertolini:2019out,Dekens:2021bro,RAMSEYMUSOLF2021136136}, leptoquark models \cite{Dekens:2018bci,Fuyuto:2018scm}, or scenarios with new light particles, such as axions \cite{deVries:2021sxz,Dekens:2022gha}.

At low-energy, flavor diagonal CPV can be described by  effective operators invariant under color and electromagnetic gauge transformations. The minimal set involves a dimension-4 operator, the QCD $\bar\theta$ term \cite{tHooft:1976rip,tHooft:1976snw}, 
the dimension-5 quark electric and chromo-electric dipole moments
(which originate from dimension-6 operators in the SMEFT)
and several dimension-6  operators, including the Weinberg three-gluon operators and four-fermion operators \cite{Pospelov:2005pr,deVries:2012ab,Jenkins:2017jig,Dekens:2019ept}.
Computing the nEDM as a function of these quark level operators is a highly non trivial task, which requires nonperturbative techniques.
In the case of the $\bar\theta$ term,
chiral techniques allow to estimate the leading pion-range contribution to the nEDM 
\cite{Crewther:1979pi}, but the presence of an unknown short-range piece at the same order induces a substantial, $\mathcal O(100\,\%)$, uncertainty. Similar chiral techniques can be used for dimension-6 operators,  but have limited predictive power
\cite{deVries:2010ah}. For these sources, estimates using  QCD sum rules \cite{Pospelov:2005pr,Haisch:2019bml}
or the quark model \cite{Yamanaka:2020kjo}
are still the state of the art, but are affected by large, not fully quantified, uncertainties.
Since the last Long Range Plan, the Lattice QCD (LQCD) community 
has invested considerable resources 
to provide nEDM calculations with reliable errors \cite{Abramczyk:2017oxr,Bhattacharya:2021lol,Bhattacharya:2022whc,Dragos:2019oxn,Alexandrou:2020mds,Liu2023}.
Several calculations of the nucleon EDM induced by the QCD $\bar\theta$ term have appeared \cite{Dragos:2019oxn,Alexandrou:2020mds,Bhattacharya:2021lol,Liu2023}. These calculations turned out to be extremely challenging, because of the small signal, which gets even smaller as the quark masses are decreased towards their physical values, and because of sizable
lattice artifacts, as for example the
contamination from nucleon-pion excited states \cite{Bhattacharya:2021lol}. 
The two calculations at the physical point yield a neutron EDM compatible with zero. The uncertainty on these calculations, however, is approaching the range of values expected if the
``chiral logarithm'' identified in Ref. \cite{Crewther:1979pi} dominates the nEDM.    
Ref. \cite{Dragos:2019oxn} used larger pion masses, $m_\pi > 410$ MeV, and it observed a signal at the $2$-$3\sigma$ level. Extrapolating to the physical point, they find a non-zero nEDM at $2\sigma$, also compatible with Ref. \cite{Crewther:1979pi}.

In the case of dimension-5 and dimension-6 operators, a further complication is the involved mixing structure of higher-dimensional operators on the lattice. Since the last LRP, the matching between the $\overline{\rm MS}$
scheme and schemes that can be implemented on the lattice has been worked out for the quark chromoelectric dipole moment and the Weinberg three-gluon operator  
\cite{Constantinou:2015ela,Bhattacharya:2015rsa,Cirigliano:2020msr,Rizik:2020naq,Kim:2021qae,Mereghetti:2021nkt}. Some work remains to be done for the Weinberg operator in the gradient flow, and for four-fermion operators.
Concerning the calculations of lattice matrix elements,
the contribution of the $u$ and $d$ quark EDMs to the neutron EDM have been determined with 8\,\% and 4\,\% uncertainties
\cite{Bhattacharya:2015esa,Gupta:2018qil}. Preliminary calculations for the chromo-electric dipole moment and the Weinberg operator also exist  \cite{Abramczyk:2017oxr,Bhattacharya:2022whc},
but they still do not have full control over all systematics. For more details, we refer to
Ref. \cite{Alarcon:2022ero}.

Building on these extremely promising preliminary results, the primary goal of the EDM LQCD effort in the next LRP will be to produce controlled   
calculations for the neutron and proton EDMs induced by the QCD $\bar\theta$ term, and by the quark and gluon chromo-EDM operators, and to start the study of four-quark operators. Moving beyond single nucleon EDMs, 
the contribution of semileptonic CPV operators to atomic and molecular EDMs is mediated by the nucleon scalar, pseudoscalar and tensor form factors \cite{Chupp:2017rkp}, which are precisely computed on the lattice \cite{Aoki2021}.
LQCD can play an important role in the determination of CPV pion-nucleon couplings \cite{deVries:2016jox},
and, once two-nucleon techniques are mature, CPV couplings in the nucleon-nucleon sector, necessary to make contact with EDMs of light ions  \cite{Song:2012yh,deVries:2020loy,Yang:2020ges}
and nuclear Schiff moments \cite{Chupp:2017rkp,Engel:2013lsa}.

A further step towards a solution of the ``inverse problem" lies in understanding the 
complementarity between measurements or bounds on the nEDM and
atomic and molecular EDMs. 
Good progress has been achieved in 
the calculation of EDMs of light nuclei,
which can be carried out using \textit{ab initio} methods
\cite{Stetcu:2008vt,deVries:2011an,Song:2012yh,Bsaisou:2014zwa,Yamanaka:2015qfa,Gnech:2019dod,Yang:2020ges,Froese:2021civ,Yamanaka:2019vec}.
Atomic EDMs, such as $^{199}$Hg,  $^{129}$Xe and $^{225}$Ra,  are, on the other  hand, affected by large theoretical 
uncertainties, due to the complicated nuclear structure entering nuclear Schiff moments.
In the last few years 
there have been new calculations for $^{199}$Hg and $^{225}$Ra \cite{Engel:2013lsa,Dobaczewski:2018nim,Yanase:2020agg}. The great progress in the application of \textit{ab initio} techniques to medium mass and heavy nuclei promises the first 
\textit{ab initio} calculation of Schiff moments in the near future
\cite{Alarcon:2022ero}. 

Finally, more theoretical work is required for a deeper understanding of the
connection between EDMs and weak scale baryogenesis. 
Open questions exist in two main areas:
(i) The study of the electroweak phase transition: here it is necessary to identify scenarios that admit a first order phase transition
or sufficiently sharp crossover
(needed to provide sufficient departure from equilibrium) and study their falsifiable signatures at the Large Hadron Collider and possible future colliders~\cite{Ramsey-Musolf:2019lsf,Wang:2022dkz}.
(ii) The generation of CP asymmetries at the phase boundary through CP-violating particle transport: 
this requires identifying and solving an appropriate set of quantum kinetic equations~\cite{Cirigliano:2009yt,Cirigliano:2011di} (QKEs), 
to track both the coherent evolution necessary for CP violating phases to manifest themselves, as well as the incoherent interactions of particles with the thermal bath. The main challenge here concerns a systematic field-theoretic formulation of QKEs for massive fermions that mix through the Higgs vacuum expectation value(s) and an efficient computational scheme to obtain numerical solutions and scan the parameter space. 

\vskip .2in
\noindent{\bf nEDM@SNS Experiment}

The nEDM@SNS experiment is the most ambitious of the new nEDM experiments, with a sensitivity two orders-of-magnitude below the present best limit. It uses the Fundamental Neutron Physics Beamline (FNPB) at the Spallation Neutron Source (SNS), where a nearly mono-energetic cold neutron beam scatters from phonons in superfluid ${}^4$He to produce UCN in the measurement cells. This allows for a relatively high density of UCN to be produced without transport losses. The superfluid ${}^4$He also acts as an electrical insulator allowing electric fields of at least 75 kV/cm as demonstrated in a medium-scale prototype system~\cite{Ito:2015}. Polarized ${}^3$He is then used as both a co-magnetometer and monitor of the UCN precession frequency.
Magnetometry is possible via SQUID sensors that measure the time-dependent magnetization of the polarized ${}^3$He, while the UCN frequency is monitored via the spin-dependent n-${}^3$He capture reaction that produces scintillation light from the reaction products. The polarized ${}^3$He not only allows for two independent techniques to be used for the EDM search (monitoring the frequency of the free precession and using critical spin dressing - see refs.~\cite{Golub_Lamoreaux_1994,Ahmed_2019}), it provides direct access to characterize one of the largest systematic effects in nEDM experiments - the so-called geometric phase effect. A small change in the operating temperature of the experiment of $\sim$~0.1~K can greatly increase the size of this false-EDM effect in ${}^3$He and thus measure the magnitude of this systematic effect in a tiny fraction of the overall experimental running time. Due to the live and in-situ neutron spin analysis, the experiment is also sensitive to time-varying axion-fields inducing EDMs with a reach to a very high axion mass relative to other nEDM experiments~\cite{Abel_2017} and even AMO EDM experiments~\cite{Roussy_2021}.


At the time of the last LRP, the experiment was beginning an intense R\&D program (Critical Component Demonstration) whereby high-fidelity prototypes of the most challenging components were constructed. In some cases, these were the full-scale components to be used in the experiment, while others demonstrated the feasibility of the techniques. At present the Magnetic Field System is being reassembled and commissioned at the SNS while the Central Detector System and Polarized ${}^3$He System are under construction. Construction of the new building to house the experiment and installation of cold neutron guides, followed by commissioning and data-taking is planned on a five year timescale. 

\vskip .2in
\noindent{\bf LANL nEDM Experiment}

The LANL nEDM experiment is based on the proven Ramsey's separated oscillatory field method at room temperature, featuring the double precession chamber geometry. 
This is an effort complementary to the nEDM@SNS. 
It takes advantage of the LANL UCN source, one of the strongest UCN sources in the world and the only operational UCN source in North America, providing the US nEDM community with an opportunity to perform an nEDM experiment and obtain competitive physics results in a shorter time scale while the development and construction of the nEDM@SNS continues. 

Soon after the last LRP, the LANL UCN source went through a major upgrade, increasing the output by a factor of 4~\cite{Ito2018}, which immediately enabled improved precision in the neutron lifetime measurement in the UCN$\tau$ experiment. A separate dedicated UCN beamline was constructed for the LANL nEDM experiment. A sufficient UCN density for an nEDM experiment with a one standard deviation sensitivity of $\sigma( d_n) = 3\times 10^{-27}$~$e\cdot$cm was demonstrated under a condition relevant for an nEDM experiment~\cite{Ito2018,Wong:2022ixk}. Development of the apparatus for the LANL nEDM experiment has been funded by LANL LDRD funds and NSF MRI. A large, high-shielding factor magnetically shielded room has been installed in the experimental area. A shielding factor of $10^5$ at 0.01~Hz and residual magnetic fields of $\lesssim 0.5~$~nT have been demonstrated. The B$_0$ coil system, the coil system to provide the uniform and stable magnetic field has been fabricated and installed inside the MSR and its performance has been characterized. Various magnetometers (a $^{199}$Hg based co-magnetometer and a $^{199}$Hg based external magnetometer as well as OPM) are being developed.  The precession chambers, electrodes, UCN valves are currently being assembled. The instruments are being commissioned for imminent data-taking with Ramsey precession measurement in FY2023. 

Various capabilities and expertise developed for the LANL nEDM experiment have benefited and are benefiting the nEDM@SNS experiment. These include: (i)~a system to scan parts for magnetic impurities and (ii)~a method to fabricate various parts that minimizes magnetic contamination.  The LANL nEDM experiment naturally provides a training ground for the next generation of scientists needed to operate the nEDM@SNS experiment.

\vskip .2in
\noindent {\bf NOPTREX}  

Neutron interactions with heavy nuclei at certain compound nuclear p-wave resonances can be used to search for T-odd interactions with high sensitivity. T-odd interactions from new sources beyond the SM can generate two types of terms in the neutron forward scattering amplitude: a P-odd/T-odd term of the form $\vec{s}_{n} \cdot (\vec{k}_n \times \vec{I})$, where $\vec{s}_{n}$ is the spin of the neutron, $\vec{k}_{n}$ is the neutron momentum, and $\vec{I}$ is the polarization of the nucleus, and a P-even/T-odd term of the form $(\vec{k}_n \cdot \vec{I})((\vec{s}_{n} \cdot (\vec{k}_n \times \vec{I}))$. These two flavors of T violation come from very different types of BSM interactions. In forward transmission experiments one can realize a null test for T which, like electric dipole moment searches, is in principle free from the effects of final state interactions~\cite{Gudkov:1990tb,Gudkov:2013dp,Bowman:2014fca}. Amplifications of $P$-odd neutron amplitudes in compound nuclear resonances by factors of $~10^6$ above the $~10^{-7}$ effects expected for weak NN amplitudes compared to strong NN amplitudes have already been observed~\cite{Mitchell:1999zz} in measurements of the P-odd longitudinal transmission asymmetry $\Delta \sigma_{P}$ in several heavy nuclei. This amplification from mixing of nearby s and p-wave resonances was predicted theoretically~\cite{Sushkov:1982fa,Bunakov:1982is} before it was measured. A similar resonance mechanism can amplify a $P$-even and $T$-odd amplitude by a factor of $10^{3}$~\cite{Barabanov:1986sz, Bunakov:1988eb, Gudkov:1991qc}. Direct experimental upper bounds on $P$-even and $T$-odd NN amplitudes~\cite{Huffman:1996ix} are only ~1\,\% of strong NN amplitudes.  

Recent (n, $\gamma$) spectroscopy measurements at JPARC~\cite{Okudaira2018} have confirmed that the 0.7 eV p-wave resonance in $^{139}$La is the best known candidate resonance for a P-odd/T-odd search. A dynamic nuclear polarized lanthanum aluminate target~\cite{Takahashi:1993np,Hautle:2000} is under development at RCNP in Japan. High performance neutron spin filters based on polarized $^{3}$He can now operate with high efficiency in the eV neutron energy range. Ideas to use n-A resonances to improve sensitivity to $P$-even and $T$-odd NN interactions by $10^{2}-10^{3}$ with tensor-aligned cryogenic targets~\cite{Barabanov:2005tc, Skoy:2007} have become more practical. Several measurements and analyses to determine the spectroscopic parameters needed to quantify the sensitivity of T-odd searches in other nuclei are underway. More complete theoretical treatments of polarized neutron optics in the resonance regime in the presence of polarized and aligned nuclear targets have appeared~\cite{Gudkov:2017sye, Gudkov:kappa, Gudkov:2020}. A proposal to JPARC for a dedicated beamline for NOPTREX is under review.    

NOPTREX is a global collaboration with a membership of more than 100 researchers from North America, Europe, and Asia which has coordinated several eV neutron spectroscopy experiments at 4 different neutron sources. 


\vskip .2in
\noindent {\bf Time Reversal Violation in Neutron Beta Decay}  

Beta decay possesses a diverse set of observables involving the spin and momenta of the final state particles.  Correlations among these can be used as stringent tests of the SM including sensitive tests of time reversal symmetry. 
Examples include the angular correlation coefficient $D$ proportional to $\vec{J} \cdot (\frac{\vec{p_{e}}}{E_e} \times \frac{\vec{p_{\nu}}}{E_\nu})$ and the angular correlation coefficient $R$ proportional to $\vec{J} \cdot (\vec{s_{e}} \times \frac{\vec{p_{e}}}{E_e})$.    Complicating the situation slightly is that $D$ or $R$ are not direct null tests for T because of contributions from T-even interactions between the final state particles, so called final state interactions (FSI). However for the neutron SM FSI can be small, 
and are calculable via heavy baryon effective field theory to the precision needed by future experiments.  SM time-reversal violation (TRV) in beta decay arises from the CKM phase 
and is so strongly suppressed~\cite{Herczeg1997TimereversalVI} that it is out of reach for foreseeable future experiments, thereby making T violation experiments in beta decay clean searches for BSM. New physics however can interfere at tree level and is only suppressed by the mass scale of the new interaction. Additional phases are generically present in BSM physics and there is no reason to think they should be small. Thus TRV tests in beta decay can potentially access physics at scales above that of accelerators.  
The current upper limit is $D=[0.94 \pm 1.89_{\text{stat}} \pm 0.97_{\text{sys}} ] \times 10^{-4}$ from the emiT collaboration~\cite{Mumm2011NewLO, PhysRevC.86.035505}. The final state interactions for $D$ are an order of magnitude smaller than the precision reached by emiT~\cite{ANDO2009109} and can be evaluated to 1\,\% precision, leaving at least two orders of magnitude of $D$ parameter space for discovery of new T violation physics. The intense new NG-C slow neutron beam at NIST provides an unique opportunity to improve limits on $D$ in a new version of an emiT-type apparatus by about a factor of 5-10, with future development of calorimetric particle detection extending the sensitivity by a factor of 40. At these levels of sensitivity one will measure FSI effects.  Importantly, such an experiment can also yield constraints on specific BSM physics that can evade limits from EDMs, for example Left-Right Symmetric Models (LRSM)~\cite{RAMSEYMUSOLF2021136136} as well as right handed neutrino couplings~\cite{ELMENOUFI201762}.  Work toward this goal has begun, with development of the necessary detector technology underway.


One can also use neutrons to search for a T-odd correlation in radiative beta decay $\hat{k_{\gamma}} \cdot (\vec{p_{e}} \times \vec{p_{\nu}})$.
This P-odd and T-odd correlation can come from a Chern-Simons contact interaction~\cite{PhysRevLett.99.261601, PhysRevD.77.085017} term which appears at N2LO order in a chiral EFT involving pions, nucleons, and electroweak fields~\cite{PhysRevD.81.013008}, while needing a new hidden strongly interacting sector to contribute to $\beta$ decay.
This correlation is sensitive to imaginary parts of the interference with the weak vector current. 
As it is spin-independent, it gives a different view into T violation than what is probed by electric dipole moments \cite{PhysRevD.86.016003,Dekens:2015usa}.

Finally, one can consider T-odd P-even correlations in beta decay.  These may prove important in disentangling radiative effects and the possible BSM physics generating an observed EDM.  TRV effects can be significantly enhanced in isospin suppressed nuclear beta decay, however existing limits are still fairly poor.

\vskip .2in

\subsection{Tests of CKM non-unitarity and other BSM physics}
\subsubsection{Motivation}
A renewed interest arose in the high-precision test of the SM prediction of the first-row CKM unitarity $|V_{ud}|^2+|V_{us}|^2+|V_{ub}|^2=1$ (though $|V_{ub}|^2$ is negligible at the current precision level). Since late 2018, Seng \textit{et al.} revisited the so-called inner radiative correction to the neutron and nuclear beta-decay, which is one of the major sources of theory uncertainties in $|V_{ud}|$, based on a novel dispersion relation analysis~\cite{Seng:2018yzq}. Using available data from neutrino-nucleus scattering, they obtained a value of the inner radiative correction significantly larger than the previous state-of-the-art value~\cite{Marciano:2005ec} with reduced uncertainty, which resulted in a reduction of the $|V_{ud}|$ central value. This finding was confirmed by several independent studies~\cite{Czarnecki:2019mwq,Seng:2020wjq,Shiells:2020fqp,Hayen:2020cxh}.

\begin{figure}[t]
	\begin{centering}
		\includegraphics[width=0.4\linewidth]{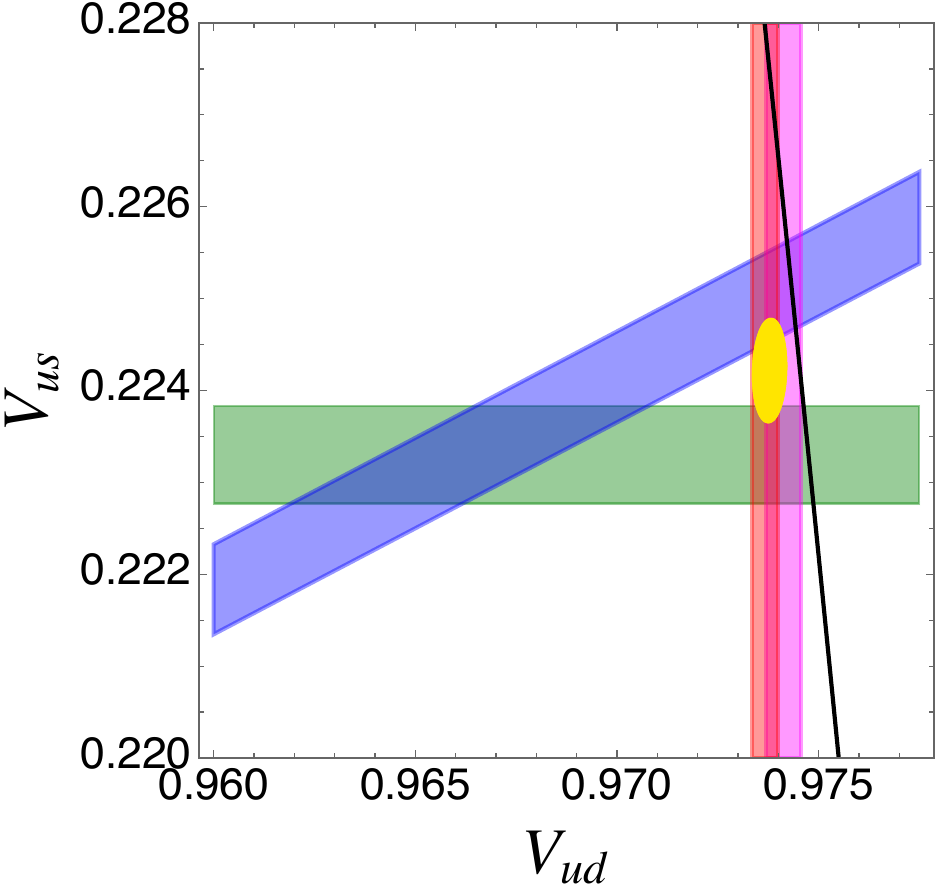}
		\par\end{centering}
	\caption{Values of $|V_{ud}|$ obtained from superallowed $0^+\rightarrow 0^+$ nuclear beta decays (red) and neutron beta decay (violet), $|V_{us}|$ from semileptonic kaon decays ($K_{\ell 3}$, green), and $|V_{us}/V_{ud}|$ from leptonic kaon/pion decays ($K_{\mu 2}/\pi_{\mu 2}$, blue). The yellow ellipse represents a global fit of the two matrix elements, and the black line assumes the first-row CKM unitarity. Figure courtesy of Vincenzo Cirigliano.~\cite{Cirigliano:2022yyo}
\label{fig:VudVus}
	}
\end{figure}

The shift of $|V_{ud}|$ has a profound impact on the precision test of the SM at low energies.
Fig.\ref{fig:VudVus} summarizes the current status of the most precisely determined values of $|V_{ud}|$, $|V_{us}|$ and $|V_{us}|/|V_{ud}|$, and several anomalies can be observed from the diagram. For instance, the combination $|V_{ud}|_{0^+}^2+|V_{us}|_{K_{\ell 3}}^2-1=-0.0021(7)$ exhibits a deficit from unitarity at the level of 3$\sigma$, and the two different determinations of $|V_{us}|$ from semileptonic and leptonic kaon decays also show a $\sim 3\sigma$ disagreement. They are now known collectively as the ``Cabibbo angle anomaly'', which provides interesting hints of BSM physics. The most precise determination of $|V_{ud}|$ presently comes from superallowed $0^+\rightarrow 0^+$ nuclear $\beta$ decays, $|V_{ud}|_{0^+}=0.97367(30)_{\text{th}}(11)_{\text{exp}}$~\cite{Cirigliano:2022yyo}, with the major uncertainty coming from nuclear structure effects~\cite{Seng:2018qru,Gorchtein:2018fxl}.
On the other hand, free neutron decay is theoretically much cleaner, but is limited by experimental uncertainties of the neutron lifetime $\tau_n$ and the axial-to-vector coupling ratio $\lambda=g_A/g_V$. Using the PDG averages, one obtains $|V_{ud}|_n^{\text{PDG}}=0.97441(13)_{\text{th}}(87)_{\text{exp}}$; however, adopting the single best measurement of $\tau_n$ from UCN$\tau$~\cite{UCNt:2021pcg} and $\lambda$ from PERKEO III~\cite{Markisch:2018ndu} respectively returns $|V_{ud}|_n^{\text{best}}=0.97413(13)_{\text{th}}(40)_{\text{exp}}$, with the total uncertainty already comparable to that from superallowed $\beta$ decays. With future improvements in the experimental precision of these quantities, neutron $\beta$ decay could eventually surpass $0^+\rightarrow 0^+$ as the best avenue to extract $|V_{ud}|$.

Finally, a comparison of an experimentally obtained value for the axial-to-vector coupling ratio, $\lambda = g_A/g_V$, from neutron $\beta$ decay and direct computation from lattice QCD is an extremely sensitive channel for probing right-handed currents \cite{Alioli:2017ces}. This was made possible by a significant increase in precision from lattice QCD determinations of $g_A$ over the last 4 years \cite{Aoki2021}, with individual calculations claiming sub-percent precision \cite{Chang2018, Walker-Loud2020}.
As attractive solutions to the Cabibbo angle anomaly propose the existence of right-handed currents \cite{Cirigliano2022b}, neutron $\beta$ decay can provide independent constraining power on both CKM unitarity and BSM physics scenarios.

BSM scenarios with charged scalars or leptoquarks can induce scalar, pseudoscalar or tensor charged-currents. Pseudoscalar couplings to electrons are severely limited by the decay $\pi \rightarrow e \nu$, which is helicity suppressed in the SM. 
The ratio $\Gamma(\pi \rightarrow e \nu)/\Gamma(\pi \rightarrow \mu \nu)$ constrains pseudoscalar currents to be five orders of magnitude weaker than SM currents \cite{Cirigliano:2012ab,Cirigliano:2013xha}, 
leaving little room for pseudoscalar contributions to other observables, such as the neutron Fierz interference term \cite{Gonzales-Alonso:2014}.
Scalar and tensor interactions can be probed quite competitively in  $\beta$ decays
and from analyzing the high transverse mass tail of the charged-current Drell-Yan process at the LHC \cite{Cirigliano:2012ab,Alioli:2018ljm,Boughezal:2021tih,Allwicher:2022gkm,Allwicher:2022mcg}. By carrying out a comprehensive analysis of $\beta$ decay data, including superallowed $\beta$ decays, neutron decay, mirror decays
and decay correlations in selected nuclei,
Falkowski, Gonz\'alez-Alonso and Naviliat-Cuncic
found bounds on the scalar and tensor interactions to be: 
$\eps_S = \left( 0.1 \pm 1.0\right) \cdot 10^{-3}$ and 
$\eps_T = \left( 0.5 \pm 1.3\right) \cdot 10^{-3}$, where $\eps_S$ 
and $\eps_T$ parametrize scalar and tensor operators in the normalization of Ref.
\cite{Falkowski:2020pma}, evaluated at the renormalization scale of $2$ GeV. Tensor interactions receive similar constraints from  pion radiative decays \cite{Gonzalez-Alonso:2018omy}.
Using the latest high-transverse-mass Drell-Yan dataset from the ATLAS collaboration
\cite{ATLAS:2019lsy}, and assuming only dimension-6 operators to contribute, we find 
$|\eps_S| < 1.1 \cdot 10^{-3}$ and 
$|\eps_T| < 1.0 \cdot 10^{-3}$  (95\,\% CL), highlighting the complementarity between these two sets of observables. 
The bounds on $\eps_T$ can be converted in a neutron Fierz interference term $|b| \lesssim 1.3 \cdot 10^{-3}$ (95\,\% CL), within reach of the next generation of experiments. 

\subsubsection{Theory prospects}

Further efforts from high-precision SM theory calculations are required to confirm the unitarity discrepancy and improve the sensitivity of neutron and nuclear $\beta$ decays to BSM physics; this requires large-scale collaborations between theorists at high- and low-energy physics, as well as interplay with experimentalists in the design of new experiments. 

In the $|V_{ud}|$ extraction, there are on-going efforts to compute the single-nucleon axial $\gamma W$-box diagram using lattice QCD, which may fully pin down the inner radiative corrections in the nucleon sector.  A proof-of-principle study based on lattice computations of four-point correlation functions was successful on the simpler pion system~\cite{Feng:2020zdc}, but to extend the method to the neutron requires more computational resources and independent studies from multiple lattice groups for cross-checking. Alternative approaches, e.g. using the Feynman-Hellmann theorem, are also possible~\cite{Seng:2019plg}.
In the nuclear sector, the nuclear structure correction $\delta_{\text{NS}}$ that currently bears the largest uncertainty in $|V_{ud}|_{0^+}$ can be rigorously expressed in terms of the difference between the nuclear and the nucleon axial $\gamma W$-box diagram~\cite{Seng:2022cnq}; the former can be computed using different ab-initio methods such as Quantum Monte Carlo (QMC), No-Core Shell Model (NCSM), Coupled Cluster (CC), In-Medium Similarity Renormalization Group (IMSRG) and Nuclear Lattice Effective Field Theory (NLEFT), depends on the mass number $A$. The ab-initio calculation of $\delta_{\text{NS}}$ for $^{10}$C $\rightarrow$ $^{10}$B may serve as the first, important prototype of such studies. Meanwhile, experimental measurements of electroweak nuclear radii may help to sort out hidden systematic errors in the isospin breaking correction to the Fermi matrix element~\cite{Seng:2022epj}, and US facilities such as FRIB can play an important role in this aspect. 

A $0.1\,\%-0.2\,\%$ precision for lattice calculation of the isospin-symmetric $g_A$ may be expected in a $\sim$ 5 year time scale, which approaches the current experimental precision of $\lambda$~\cite{WalkerLoud}. However, in order to compare with experimental measurements, one needs to understand the radiative corrections to $g_A$. Unlike $g_V$, the $\gamma W$-box diagram contribution to $g_A$ is well under control~\cite{Hayen:2020cxh,Gorchtein2021}, however it was recently pointed out that a much larger contribution comes from the vertex correction to the neutron charged weak form factor associated to the pion mass splitting~\cite{Cirigliano:2022hob}. An appropriate combination of effective field theory (EFT) and the classical Sirlin representation of radiative corrections~\cite{Sirlin:1977sv,Seng:2019lxf,Seng:2021syx} may help to express such correction in terms of well-defined hadronic matrix elements which permit a direct lattice calculation, opening the channel for stringent right-handed current searches.

\subsubsection{International context}
Meson decay measurements will also provide some input to the Cabbibo-anomaly problem during the coming LRP period.  Although not a critical limiter on the precision of $V_{us}$, new high precision kaon branching ratios should be available from the NA62 experiment.  Given that the current branching ratio for $K\mu 2$ comes essentially only from the KLOE experiment, these results may have an impact \cite{Cirigliano:2022yyo}.  An improved determination of $V_{us}$ from an inclusive measurement of $\tau$-lepton decays may also be possible from the Belle II experiment, although challenges exist for this measurement both on the experimental and theoretical sides. A proposed experiment at PSI, PIONEER (\cite{PIONEER2022}) aims to start taking pion decay data in 2029, with a first goal to improve the precision with which lepton flavor universality is known by an order of magnitude. The follow up (after the upcoming LRP period) would be an improved measurement of the branching ratio for $\pi ^+ \rightarrow \pi ^0 e^+ \nu$ for an independent determination of $V_{ud}$. This would come in two stages, with the first increment an improvement in precision of this branching ratio by a factor of three over the PIBETA experiment \cite{Pocanic2004}, to establish a ratio of decay rates for the $Kl3$ decays which is less sensitive to systematic errors from form factor calculations, similar to the ratio of $Kl2$ to $\pi l2$ decays.  This first phase would then provide necessary input to a second phase with the potential to reach the precision of the superallowed decays.
\begin{figure}[t]
	\begin{centering}
		\includegraphics[width=0.8\linewidth]{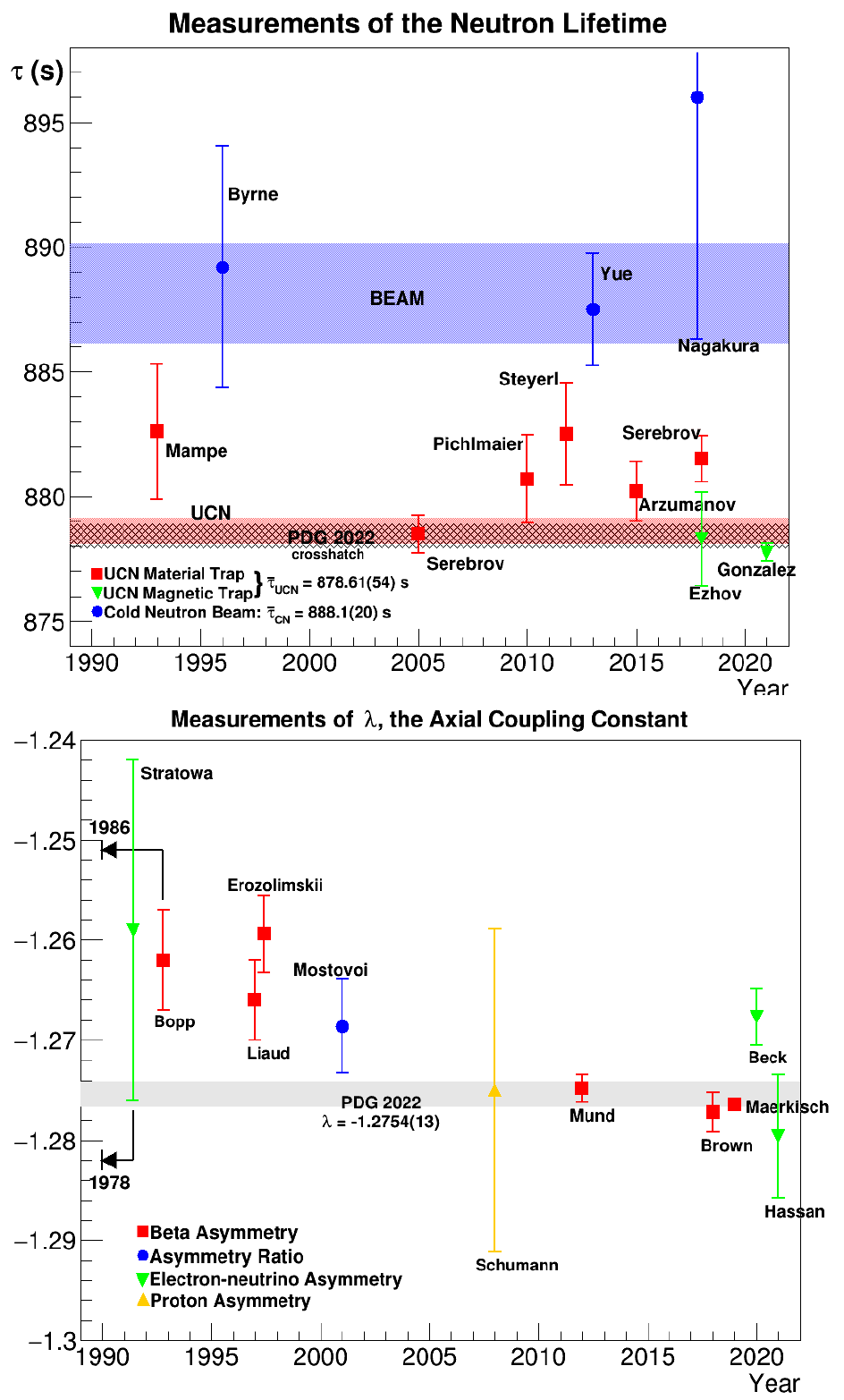}
		\par\end{centering}
  \caption{Top-most recent and/or precise lifetime results from the global neutron beta decay experimental program, including measurements using UCN (squares, with ${\bar \tau}_{UCN}=878.6(5)$~s in pink band with uncertainty scaled by 1.9)~\cite{Mampe:1993a,Serebrov2005a,Pichlmaier:2010zz,Steyerl:2012a,Arzumanov2015,Serebrov2018,Ezhov2018,Gonzalez2021}
  and cold neutron beams (circles, ${\bar \tau}_{UCN}=888.1(2.0)$~s in blue
  band)~\cite{Byrne:1996a,Yue:2013a,Nagakura:2017a} and the crosshatch band indicating the PDG 2022
  average~\cite{PDG2022}, $\bar \tau _{PDG} = 878.4(5)$~s with scale factor 1.8.  Bottom -- most recent and/or precise measurements of $\lambda=g_A/g_V$, including measurements using the $\beta$-asymmetry (squares)~\cite{bopp86,yerozolimsky97,liaud97,Mund:2012a,Brown2018,Maerkisch:2018a}, ratios
  of electron-proton coincidence asymmetries (circles)~\cite{Mostovoi2001}, the proton-asymmetry
  (triangles)~\cite{Schumann:2008b} and the electron-antineutrino asymmetry (inverted
  triangles)~\cite{Stratowa1978,Byrne:1996a,Beck:2020a,Hassan:2021}.
  Also shown is the PDG 2022 average~\cite{PDG2022} $\lambda = -1.2754(13)$ (scale factor $2.7$). 
 }
\label{fig:betadecay_world}
\end{figure}

Over the last LRP period there has been a burst of productivity from the neutron beta-decay experimental community (see Fig. \ref{fig:betadecay_world}), with four new experimental results for the neutron lifetime and four new (or updated) results for angular correlations.  The lifetime measurements include two experiments reporting magnetically trapped ultracold neutrons -- the experiments of Ezhov {\it et al}~\cite{Ezhov2018} and UCN$\tau$~\cite{Gonzalez2021}, with magnetic trapping now providing the standard for UCN storage experiments.  The ground-breaking experiment of Serebrov {\it et al}~\cite{Serebrov2005a} remains the most precise measurement of the lifetime using a material trap. The team led by Serebrov carried out a revised experiment using a similar approach; their result was published in 2018~\cite{Serebrov2018}. Furthermore, Arzumanov {\it et al} upgraded their previous experiment~\cite{Arzumanov2012} and published a new result in 2015~ \cite{Arzumanov2015}. 

In addition to on-going work on beam-based lifetime measurements at NIST (discussed below), steady progress has been made by the J-PARC beam experiment~\cite{Ichikawa:2022ZH,Mogi:2022ov}, with uncertainties limited by systematic effects at about the 7~s level now limiting the precision, and more running and refinement expected in the next few years. Furthermore, an innovative determination of the lifetime from the dependence of the number of thermal neutrons as a function of altitude above the Moon's surface, with that surface being the source of thermalized neutrons, has reached a 15~s precision~\cite{wilson_ref}.

The long-standing discrepancy between the reported value for the neutron lifetime in storage and beam experiments remains unresolved, and has led to a flurry of proposals and activity to find additional neutron beta decay channels \cite{Fornal2018,Fornal2020,Dubbers2019,Berezhiani2009,Broussard2022}, with no positive results so far.

There were four new angular correlation measurements: the $\beta -{\bar \nu}$ correlations reported by the aCORN experiment~\cite{Hassan:2021} and the aSPECT collaboration~\cite{Beck:2020a} and the $\beta$-asymmetry measurements reported by the UCNA collaboration\cite{Brown2018} and the PERKEO III collaboration~\cite{Maerkisch2019}.  As can be seen from Fig. \ref{fig:betadecay_world}, the results from beta asymmetry measurements are consistent, but differ by $3\sigma$ from the determination from $\beta -{\bar \nu}$ correlation in aSPECT..  The PERKEO III collaboration also published a measurement of the Fierz term in neutron beta decay~\cite{Saul2020} and limits for decays to dark particles with the emission of $e^+ - e^-$ pairs\cite{Klopf2019} (from PERKEO II data).


In the near future there are three new high-precision lifetime experiments underway at international laboratories.  Two are magnetic storage experiments: the $\tau$Spect experiment, currently based at Mainz and the PeNELOPE experiment, under development at Munich. Both now target ultimate uncertainties at the $0.1$~s level. The J-PARC cold neutron beam experiment has as its goal a precision of 1~s, with a possible upgrade to a magnetically-guided experiment also under development.
For angular correlations, there are two experiments underway at international laboratories, with a measurement of the $\beta$-asymmetry with PERC targeting 0.05\,\% precision at the Heinz Maier-Leibnitz Zentrum in Munich, with a later upgrade planned for the ANNI beamline at ESS~\cite{ANNI2019,ESS2022}, and BRAND, still in a prototyping phase and aiming for an precision at the 0.1\,\% level at the ESS for a large set of angular correlations (with running on a time scale of about 2030 at ESS). 

\subsubsection{Progress}
  One of the most significant steps in the field was an improvement of a factor of two over previous lifetime measurements by the LANL-based UCN$\tau$ experiment, which reported a first physics result in 2018~\cite{Pattie:2018a} and a value of $877.75(22)_{stat}(+22/-16)_{sys}$~s~\cite{Gonzalez2021} in 2021. The UCN$\tau$ experiment uses an asymmetrical, bowl-shaped magnetic trap to store neutrons. UCN are introduced from the bottom of the trap, and the spectra is prepared so that detected UCN have insufficient energy to overcome the gravitational potential barrier required to exit through the top surface of the bowl.  The stored populations are monitored using an {\it in situ} detector lowered into the trap.  The combination of extremely low UCN loss from the trap, strong control of systematic uncertainties through the {\it in situ} detector, and the large number of stored UCN possible in this high volume trap coupled to the LANL UCN source~\cite{Saunders:2013RSI,Ito2018}, have established this as the highest precision experiment to date.  The UCN$\tau$ result reinforces the ``lifetime puzzle'' and pulls the global average for the lifetime to $\tau_n = 878.4(0.5)$~s, with a scale factor of 1.8~\cite{PDG2022}, suggesting underestimated systematic uncertainties even within the UCN experiments.  The UCN$\tau$ collaboration also published the strongest direct limits to date for neutron decay to a ``dark" particle with the emission of a $\gamma$-ray~\cite{Tang2018}.

  The beam-based BL2 experiment is ongoing at the National Institute of Standards and Technology (NIST), and is designed to probe the systematic uncertainty budget of the BL1 experiment and provide an improved value for the neutron lifetime.  The BL1 experiment was last updated by Yue {\it et al.}~\cite{Yue:2013a}, which provided the driving motivation for the current ``lifetime-puzzle".  Experimental running of BL2 should be complete in 2023, with publication following on a few year time scale and a targeted uncertainty of better than 2~s.  The NIST team also published a high precision measurement of the radiative decay branch in neutron decay in 2016~\cite{Bales:2016a} which stands as the definitive measurement for that process.
 
 The UCNA experiment published a ``final" analysis of the $\beta$-asymmetry in 2018~\cite{Brown2018} with a combined result (all UCNA measurements) for the $\beta$-Asymmetry parameter of $A_0= -0.12015(34)_{stat}(63)_{syst}$, which yielded for the axial coupling constant, $\lambda = -1.2772(20)$.  UCNA is the only angular correlation experiment which has used UCN, exploiting the ability to produce and store very highly polarized populations of UCN with negligible neutron-generated backgrounds.  UCNA remains the highest precision, independent cross-check of the cold neutron beam measurements PERKEO II and PERKEO III which define the state-of-the-art determinations of $\lambda$.  The UCNA collaboration also published the most precise limits for neutron decay to dark particles with the emission of $e^+ -e^-$ pairs~\cite{Sun2018}, and the first direct limits on Fierz terms in neutron decay~\cite{Hickerson2017,Sun2020}.
 
 Rapid progress has also been made on the $\beta$-$\bar \nu$ asymmetry.  The aCORN experiment produced the first increment in precision for the $\beta$-$\bar \nu$ parameter $a$ in 15 years with their publication in 2017~\cite{Darius:2017a} and a final result in 2021 of $a = -0.10782(124)_{stat}(133)_{sys}$~\cite{Hassan:2021} and $\lambda = -1.2796(62)$.  Two other experiments are under development at NIST: a third run of aCORN and aCORN-B, a follow up experiment using the aCORN spectrometer and a polarized neutron beam to measure the $\bar \nu$-asymmetry coefficient, ``B".  Although no schedule is defined as yet, these experiments would target a goal below 1\,\% in the $a$ parameter and below 0.3\,\% in the $B$ parameter.
 
 As mentioned above, the current precision for $a$ is defined by the results of the aSPECT experiment. The Nab experiment, currently in a commissioning phase at the SNS, targets a factor of 7 improvement in the $a$ parameter relative to aSPECT. First decay data for Nab is possible before the coming shut-down at the SNS in the fall of 2023, and data-taking planned until roughly 2025, with an expected relative uncertainty for $\lambda$ of about 0.04\,\%. 
 
 
\subsubsection{Prospects}
The primary goal for the US neutron $\beta$-decay community during the next LRP is to determine the value of $V_{ud}$ from neutron decay with a precision competitive with the $0^{+}\rightarrow 0^+$ decays.  Using the Particle Data Group global averages as the standard, this will require less than a factor of two improvement in the uncertainty for the lifetime, and a factor of 3 for $\lambda$ ($\Delta\tau_n \sim 0.3$~s and $\Delta\lambda/|\lambda|\sim 0.03\,\%$ is needed, which includes understanding of potential discrepancies between methods at this level).  If the US experimental program is successful, these goals can be achieved.  The impact of this program would be a critical confirmation of the Cabbibo anomaly, and importantly, one without the nuclear corrections required for the $0^+ \rightarrow 0^+$ decays.  This high precision data will also provide improved constraints on exotic couplings (through their influence on the decay rates).

There are three US-based lifetime experiments planned for the next LRP period: UCN$\tau +$ and UCNPro$\beta$e at the LANL UCN source and BL3 at NIST.  The strategy for UCN$\tau +$ is to improve the statistical uncertainty using an adiabatic transfer technique to load the existing magnetic trap. Because a number of the constraints for key systematic uncertainties (including contributions from quasi-bound UCN and phase-space evolution) are limited by the statistical sensitivity of the experiment, uncertainties below 0.15~s appear feasible. Commissioning and a start for running of UCN$\tau +$ is planned for 2024.  UCNPro$\beta$e is designed to measure the branching ratio for $\beta$-decay relative to all decay modes (the total disappearance rate) for neutron decay.  The sensitivity target for the branching ratio is 1.2~s, 
giving UCNPro$\beta$e the potential to play a critical role if the discrepancy between the NIST beam experiments and UCN storage experiments persists. Commissioning of UCNPro$\beta$e is planned for 2025, with final data taking in 2027. The BL3 experiment builds on the strategies developed in Yue {\it et al.}~\cite{Yue:2013a} for high precision determination of the density of the neutron beam, with a scaled-up trap volume and increased neutron flux at the NG-C beamline.  The ability to achieve 1~s precision in a day of running will ensure that extensive characterization of the systematic error budget will be possible. The initial run at NIST is planned to begin in 2026, with a precision goal $< 0.3$~s.

There are also three US-based angular correlation experiments which are already underway or could be mounted during the next LRP period which can also provide a precision for $\lambda$ comparable the most precise measurement to date, PERKEO III.  The only experiment currently in commissioning is Nab~\cite{Baes2014,Fry19}, with an expected sensitivity to $\lambda$ of about 0.04\,\%. This experiment is the first to use the combined electron and proton energy spectrum to reach the ultimate sensitivity to the $\beta$-$\bar \nu$ parameter.  This experiment has the potential to resolve the current tension between recent $\beta$-asymmetry measurements and the aSPECT result. It will also provide a critical contribution to the high precision data set, with the measurement subject to a distinctly different set of systematic uncertainties than the previous and on-going $\beta$-asymmetry measurements. A natural extension of the Nab experiment can make use of the Nab spectrometer and a polarized neutron beam to perform simultaneous measurements of the $\beta$-asymmetry and angular correlations involving polarized protons.  This experiment, called pNab, will require almost no modification of the existing Nab apparatus, since the capability for highly polarized neutron beams and spin analysis is now incorporated into the baseline capability for Nab. Although this experiment is not yet approved for the FNPB beamline, it would provide a new measurement of $\lambda$ with a goal of $\Delta\lambda/\lambda=0.02\,\%$ and new methods to control sources of systematic uncertainties through coincident detection of electrons and protons and ratios of spin-dependent observables.  

Research and development towards an upgrade of the UCNA experiment is currently underway at the Los Alamos UCN source. This experiment, called UCNA+, would utilize the high UCN densities available in the LANL UCN source to reduce statistical uncertainties and an improved detector package to minimize scattering corrections. These improvements push the projected sensitivity for the $\beta$-asymmetry below 0.2\,\%, making it comparable in sensitivity to Nab and PERKEO III. Given the current uncertainty in the schedule for Perc and the control of key systematic uncertainties through the use of UCN, UCNA+ could make a significant impact on the drive towards a global neutron $\beta$-decay data set competitive with the superallowed decays.

Direct limits on exotic couplings through measurements of the final state electron energy dependence of neutron decays hold the promise of very strong direct constraints on scalar and tensor couplings~\cite{Bhattacharya:2012a}.  Experiments of this kind are very challenging, with progress limited by the technologies currently available.  At present, the Nab experiment has as one of its goals the direct measurement of the $\beta$ energy spectrum resulting in an uncertainty of $3\cdot 10^{-3}$ in a possible Fierz interference term, limited by detector-related uncertainties.  Measurements of this kind are very sensitive to exotic couplings, with this measurement providing constraints comparable to the neutron lifetime and asymmetry data for tensor couplings.  Experimental techniques which can probe these couplings more sensitively, such as ``broadband" cyclotron emission spectroscopy~\cite{Asner:2015a,Byron2022} are under development now for nuclear $\beta$-decay fundamental symmetries experiments, and may emerge as a viable path to improved constraints during the coming LRP period.

The potential impact of the US program on the Cabbibo-anomaly is very high.  The planned neutron lifetime measurements provide a robust basis to establish the lifetime at the required 0.3~s level while accomodating typical scatter between various experimental approaches. The US is in a leadership position with these measurements.  They also provide a path to clarify the current discrepancy between beam and storage experiments.  BL3 has the precision and control of systematic uncertainty to definitively confirm or contradict the BL1 measurement.  If BL3 confirms the BL1 result, UCNPro$\beta$e is designed to directly determine the branching ratio for neutron decay via the charged weak current, potentially identifying new physics as the explanation of the lifetime puzzle.  In contrast, there is only one new angular correlation measurement, Nab, mounted at an operational beamline and being commissioned.  A successful measurement (as mentioned) can potentially unambiguously confirm or contradict the value of the $a$ parameter determined by aSPECT.  It also has sufficient precision to effectively achieve the 0.03\,\% goal in $\lambda$, when taken together with PERKEO III.  Given that Nab is the only angular correlation measurement currently scheduled to take data, the US research program does not yet incorporate at least one alternative, high precision experimental approach for a robust determination of $\lambda$.    The availability of measurements with significantly different methodology and sources of systematic error (as with the experimental program underway for the neutron lifetime) has historically been extremely important in this subfield.
There is a clear benefit to implementing pNab and/or UCNA+ to ensure the US program plays a decisive role in the evolution of our understanding of the Cabbibo-anomaly and ensuring US leadership on this problem.  Although the Perc collaboration pursues similar goals (with somewhat more optimistic precision targets for the $\lambda$ parameter), the US experimental program has the capability to provide leadership for the global neutron beta decay program, establishing the neutron as the definitive reference system for the charged weak current of the nucleon.


\subsection{Parity violations in Nuclear Systems}
\subsubsection{Motivation}

NN weak interaction amplitudes probe one of the most poorly-understood sectors of the SM. The relative sizes of different quark-quark weak interaction amplitudes, which in turn induce NN weak interactions, are very sensitive to quark-quark correlations in the nucleon and to low energy nonperturbative NN strong interaction dynamics. The measurement of NN weak amplitudes therefore offers a unique, dynamically-rich regime in which to test the standard electroweak model.  NN weak interactions provide a new opportunity to develop and test theoretical methods in low energy strong interaction theory such as effective field theory and lattice gauge theory~\cite{Wasem:2011tp} and make predictions in a challenging but calculable strongly interacting few nucleon systems~\cite{Adelberger:1985ik,Desplanques:1998ak, Ramsey-Musolf:2006vfz,Haxton:2013aca,Schindler:2013yua,deVries2013, Gardner:2017xyl, deVries:2020iea}. 

The NN weak interaction is also a test case for our ability to trace symmetry-violating effects of a known quark-quark interaction across many nonperturbative strong interaction scales. This is an exercise that also must be performed for many other searches for symmetry-violating low energy nuclear observables beyond the SM such as electric dipole moments and neutrinoless double beta decay. Interpreting such experiments requires calculation of matrix elements in heavy nuclei, which cannot be directly measured and where theoretical methods give a wide range of results.  A quantitative understanding of NN weak amplitudes, together with measurements of phenomena depending on these amplitudes in heavier nuclei, could provide useful benchmarks for the relevant aspects of nuclear structure theory. For example, NN weak interactions induce parity-odd nuclear anapole moments~\cite{Zeldovich1957,Flambaum:1980sb}, whose effect can be measured in experiments using methods from atomic/molecular/optical physics and QIS~\cite{Wood:1997zq}.  Calculations of atomic/molecular structure needed to determine anapole moments from such measurements routinely achieve uncertainties of $<10\%$~~\cite{Hao:2020zlz, Hao2018PRA}, and in some atoms much lower~\cite{Derevianko2007Proceedings}. NN weak amplitudes can also be used to test the statistical theory of symmetry violation in neutron-nucleus resonances~\cite{Tomsovic2000} against the extensive data set from the TRIPLE collaboration.




\subsubsection{Progress}

The three new NN weak interaction measurements reported since the last LRP 
are the only new precise experimental results on NN weak interactions in few body systems in several years. The NPDGamma collaboration reported~\cite{NPDGamma:2018vhh} the parity-odd asymmetry $A^{np}_{\gamma}=[-3.0 \pm 1.4(stat) \pm 0.2(sys)] \times 10^{-8}$ in $\vec{n}+p \to D + \gamma$ to determine the $\Delta I=1$, $^{3}S_{1} \to ^{3}P_{1}$ component of the weak nucleon-nucleon interaction. The n3He Collaboration reported~\cite{n3He:2020zwd} the smallest asymmetry of any parity-odd asymmetry in NN interactions measured so far: $A_{PV}=[1.58 \pm 0.97 (stat) \pm 0.24 (sys)] \times 10^{-8}$ in the emission direction of the proton in polarized neutron capture on $^{3}$He,  $\vec{n}+^{3}$He $\to ^{3}$H $+ p$. Both of these measurements were completed at the FnPB beam at SNS. The final analysis of an upper bound on parity-odd neutron rotary power in n$+^{4}$He measured at NIST of $d\phi/dz=[+2.1 \pm 8.3(stat.) \pm 2.9(sys.)] \times 10^{-7}$ rad/m~\cite{Swanson:2019cld} was published. All of these experiments required the development of new experimental techniques that were shown to limit systematic uncertainties in neutron interactions with matter at the ppb scale, well below the uncertainty from neutron counting statistics. This opens the way for improved measurements at higher intensity neutron beams. The NPDGamma result is in mild tension with previous data on the  $^{3}S_{1} \to ^{3}P_{1}$ amplitude from circular polarization measurements in $^{18}$F decay~\cite{Adelberger:1983,Page:1987}, with a theory calculation calibrated from first forbidden beta decay data~\cite{Haxton:1981}.  

The theory of NN weak interactions has undergone a qualitative change. The well-known DDH model used to guide theoretical and experimental work has been supplemented with improved input~\cite{Gardner:2022mxf} and surpassed by theory approaches with a more direct connection to QCD, such as lattice gauge theory~\cite{Wasem:2011tp}, pionless and chiral effective field theories ~\cite{Haxton:2013aca,Schindler:2013yua, deVries2013,Gardner:2017xyl}, and related ``hybrid" approaches involving combinations of lattice and EFT calculations ~\cite{Feng2018, Sen:2021dcb} with dynamical approximations using the $1/N_{c}$ expansion~\cite{Zhu2009,Phillips2015, Schindler2016} and the factorization approximation for nucleon-meson matrix elements~\cite{Gardner:2022dwi}. This work has mapped out a path toward the determination of the 5 low-energy constants in the pionless EFT NN weak interaction and has enabled specific predictions for NN weak processes under different dynamical assumptions. Parallel improvements in the theoretical treatment of strong interactions has led to more reliable predictions for the relative contributions of different NN weak amplitudes in few body systems~\cite{Viviani2014,Hyun2017,Lazauskas2019}.

A new technique to measure nuclear anapole moments of heavy nuclei~\cite{Demille2008} also has been demonstrated~\cite{Altuntas2018}. This method takes advantage of systematically small energy differences between opposite-parity levels in molecules~\cite{Flambaum1985}, which can be tuned experimentally to near-degeneracy using external magnetic fields. This greatly enhances the P-odd asymmetry, as compared to earlier experiments using atoms~\cite{Wood:1997zq}. A recent experiment demonstrated sensitivity and systematic error control sufficient to measure the anapole moments of many heavy nuclei at the $\sim$10\% level~\cite{Altuntas2018}.  In addition, for the first time calculations of anapole moments using modern methods to determine nuclear structure---here, the no-core shell model---were performed, for light nuclei~\cite{Hao:2020zlz}.

\subsubsection{Prospects}
The primary goal of this research program is to (over)determine the low energy NN weak interaction amplitudes. More experimental and theoretical work in atomic, molecular, and nuclear systems is needed to reach this goal. Continued extension of the  NN weak EFT calculations to more few body systems is essential for the interpretation of measurements and is the subject of active ongoing work. The initial goal for lattice gauge theory is to calculate the $\Delta I=2$ NN weak amplitude, which is computationally easier to access than the other NN weak amplitudes due to the absence of disconnected diagrams. Additional work on dynamical models which can help develop insight into the physics behind the relative size of different NN weak interaction amplitudes is also needed.   

The n-$^{4}$He parity-odd neutron spin rotation experiment in preparation for NIST now has a projected sensitivity for the P-odd rotary power of $10^{-8}$ rad/meter for 1/2 year of running on the NG-C beam at NIST after the cold source upgrade. This measurement can provide a strong constraint on a known linear combination of NN weak amplitudes and can distinguish between recent predictions based on $1/N_{c}$ arguments and a combined renormalization group $+$ lattice-constrained factorization calculation. The European Spallation Source will soon evaluate a proposal~\cite{ANNI2019} for a pulsed slow neutron beamline with time-averaged intensity comparable to NIST and ILL where new opportunities for NN weak interaction experiments can be realized. Two examples of possible NN weak experiments which can take special advantage of the strengths of the ESS are (1) neutron-proton parity-odd spin rotation, which is one of the few experimentally-accessible observables with sensitivity to the $\Delta I=2$ NN weak amplitude, and (2) parity-odd gamma asymmetry in $\vec{n}+D \to T+\gamma$, which is a sufficiently simple system to be treatable in terms of two-body NN weak amplitudes~\cite{Song2012}. Another observable sensitive to the $\Delta I=2$ NN weak amplitude is parity-odd photodisintegration in $\vec{\gamma}+D \to n+p$ very near threshold, which can be pursued in principle at an upgraded HiGS facility~\cite{Howell:2020nob}.

An additional major goal is to use the NN weak amplitudes determined from these few-nucleon measurements to calculate parity-odd observables in mid-mass and heavy nuclei. These will then be quantitatively compared to measurements of these observables, as a means to benchmark the uncertainties in nuclear structure techniques used for calculating matrix elements for neutrinoless double beta decay, nuclear Schiff moment, and other interesting phenomena. More experimental and theoretical work also is needed to reach this goal.

The ZOMBIES experiment projects to build on its recent proof-of-principle to measure anapole moments of several nuclei in the range $Z \gtrsim 40$~\cite{Demille2008, Altuntas2018}, initially $^{137}$Ba in the molecule BaF.  New approaches are being explored~\cite{Norrgard2019, Hutzler2020, Udrescu2022BAPS} to enable measuring anapole moments of very light nuclei, where accurate structure nuclear calculations are already being performed.  These rely on the same principle as ZOMBIES, but use recently-developed methods for increased quantum control of molecules---such as direct laser cooling \cite{Shuman2010Nature, Vilas2022Nature}, and quantum state readout of trapped molecular ions \cite{Cairncross2017PRL}---to achieve better energy resolution and enable measurements even of radioactive nuclei~\cite{GarciaRuiz2020Nature}. Several experiments aiming to measure anapole moments in atoms are also in development~\cite{Tsigutkin:2009zz, Gwinner2022QST}. In the meantime, improved calculations of anapole moments in light nuclei ($Z \lesssim 10-20$) are believed possible~\cite{Hao:2020zlz}, and promising new approaches for heavy nuclei are being pursued.

\subsection{Baryon Number Violation}

\subsubsection{Motivation and International Context}

Baryon Number Violation (BNV) is one of the key ingredients identified by Sakharov as required for the mechanism behind baryogenesis. 
Although the SM predicts BNV and Lepton Number Violation (LNV) due to electroweak instanton effects, BNV has not yet been observed experimentally, and therefore might be anticipated.
BNV processes like proton decay are now heavily constrained, and these $\mathcal{B-L}$ conserving processes don't provide a solution for baryogenesis due to sphaleron processes~\cite{Dev:2022jbf}. LNV may also indicate an explanation for the BAU via leptogenesis~\cite{Fukugita:1986hr} and is being rigorously explored by the neutrinoless double beta decay community. Neutron oscillations $n\rightarrow\bar{n}$ are $\Delta\mathcal{B}=2$ and $\mathcal{B-L}$-violating, and therefore are attractive to pursue to explain the BAU~\cite{Phillips:2014fgb}. 
Possibly related processes of $n\rightarrow{n'}$ where $n'$ belongs to a dark sector~\cite{Berezhiani:2005hv} and implications for BNV have also been discussed. 
The possibility of $n\rightarrow\bar{n}$ was originally introduced in~\cite{Kuzmin:1970nx} and aspects of models which predict neutron oscillations have been summarized in recent Snowmass whitepapers~\cite{Proceedings:2020nzz, Barrow:2022gsu, Dev:2022jbf}. The importance of searches for $n\rightarrow\bar{n}$ was highlighted in the Recent Snowmass Frontier Summary Reports~\cite{FileviezPerez:2022ypk, Artuso:2022ouk}.

Searches for 
$n\rightarrow\bar{n}$ are performed utilizing either large volume detectors or free neutrons, approaches which are complementary in technique as well as in discriminating power among different theoretical models. 
The most sensitive search for $n\rightarrow\bar{n}$ has been performed using the SuperKamiokande detector with constraints recently published of $\tau_{n\bar{n}} > 4.7\times10^8$\,s (90\,\% C.L.)~\cite{Super-Kamiokande:2020bov}. Future detectors could reach similar or better sensitivity~\cite{Dev:2022jbf}, including NOvA, MicroBooNE, Hyper-K, and DUNE, if background events and other systematic effects can be sufficiently well controlled.  
DUNE has an expected reach of $\tau_{n\bar{n}} > 5.53\times10^8$\,s (90\,\% C.L.)~\cite{DUNE:2020ypp} which could be increased further with improved modeling and analysis techniques~\cite{Barrow:2021odz}. 
In stark contrast to intranuclear searches, free searches for $n\rightarrow\bar{n}$ can be expected to be background-free, and offer the possibility of real discovery potential. 
A 
search performed using free neutrons at the ILL 
detected zero candidate events and zero background events, 
obtaining the limit $\tau_{n\bar{n}} > 0.86\times10^8$\,s (90\,\% C.L.)~\cite{BaldoCeolin:1994jz}. 
Leveraging technological developments in neutron optics and detection capabilities in the ensuing decades, 
the NNBAR 
experiment at the ESS has been designed to improve the experimental sensitivity to $n\rightarrow\bar{n}$ by three orders of magnitude compared to ILL, 
reaching $\tau_{n\bar{n}}\sim10^{9-10}\,$s~\cite{Addazi:2020nlz}. 
Constraints on the $\Delta\mathcal{B}=1$ process $n\rightarrow n'$ are much less stringent than for $n\rightarrow \bar{n}$ and best constraints have been obtained by searching for disappearance of bottled UCN~\cite{Serebrov:2008her, Altarev:2009tg, nEDM:2020ekj, Mohanmurthy:2022dbt} and fast reactor neutrons~\cite{Almazan:2021fvo, Stasser:2020jct} with some anomalous signals reported~\cite{Berezhiani:2012rq,Berezhiani:2017jkn}.  

\subsubsection{Progress and Prospects}

The ESS has accommodated critical provisions for a new high sensitivity $n\rightarrow\bar{n}$ search, including the world-unique Large Beam Port, more than 200\,m available for beamguide, and optimization of the lower moderator for fundamental physics including NNBAR~\cite{Abele:2022iml}.  
Progress toward the design of the moderator, neutron reflector, beamline, shielding, and $\bar{n}$ detector as part of the HighNESS project (with key US participation) has been reported~\cite{Backman:2022szk}. 
Recent studies have investigated how to minimize loss of sensitivity by avoiding a ``clock-reset'' after neutron reflection from different materials~\cite{Kerbikov:2018mct,Nesvizhevsky:2018tft}, with potentially interesting
consequences for the required scale of the project. 
R\&D efforts in the next decade would be very timely to explore and fully understand the impact on the final experiment sensitivity and project costs. Support for activities in the US are needed to ensure current leadership roles are not lost and that new workforce can be trained to support the future project.

To build toward the future high sensitivity 
$n\rightarrow\bar{n}$ search in NNBAR, a much smaller scale program 
of searches for $n\rightarrow n'$ 
has been initiated at ORNL, accessing complementary science of the question of the nature of dark matter. 
A first demonstration 
recently excluded~\cite{Broussard:2021eyr} the possibility that $n\rightarrow n'$ provides an explanation for the neutron lifetime puzzle as proposed in~\cite{Berezhiani:2018eds}. 
A program of searches for different mechansims of $n\rightarrow n'$ utilizing existing neutron scattering instruments within the User Program of the High Flux Isotope Reactor~\cite{Berezhiani:2017azg, Broussard:2017yev, Broussard:2019tgw} has now commenced  
with an ultimate goal of a search for $n\rightarrow n'/\bar{n}'\rightarrow \bar{n}$ later in the decade, as a shortcut to $n\rightarrow \bar{n}$ through the dark sector~\cite{Berezhiani:2020vbe}. 
This effort serves as a staged program with requiring only modest, incremental investment, and 
represents a unique opportunity to develop US expertise and provide  
technical development for the 
large scale O(\$100M) NNBAR project with an extremely economical entry-point. 
While the NNBAR project is beyond this Long Range Plan period, 
a modest investment in this decade can ensure that an exciting opportunity to explore baryon number violation with exceptional sensitivity is not missed.


\subsection{Other Precision Measurements}

The possible existence of new interactions in nature with sub-millimeter ranges, corresponding to exchange boson masses above 1 meV and with very weak couplings to matter has been discussed for some time~\cite{Leitner,Hill}. Particles which might mediate such interactions started to be referred to generically as WISPs (Weakly-Interacting sub-eV Particles) about a decade ago~\cite{Jae10}. The extended symmetries present in many theories beyond the SM, including string theories, are typically broken at some high energy scale, leading to weakly-coupled light particles with relatively long-range interactions~\cite{Arvanitaki2010, PDG14}. A continuous chiral symmetry spontaneously broken at some scale $M$ generates a massless pseudoscalar mode which couples to massive fermions $m$ with a coupling of order $g=m/M$. When the symmetry is also explicitly broken at scale $\Lambda$, the mode can become a pseudo-Goldstone boson of order $m_{boson}=\Lambda^{2}/M$~\cite{Weinberg72}.

One can conduct a reasonably general classification of interactions between nonrelativistic spin $1/2$ fermions assuming the usual constraints from relativity and quantum mechanics, which for the weakly-coupled interactions of interest leads to the usual exchange boson interaction mechanism, supplemented with (in general spin dependent) couplings at the vertices. Most descriptions have included spin $0$ or spin $1$ boson exchange~\cite{Dob06, Fadeev2019b}. Later on this work developed an interesting overlap with a set of dark matter models which also could be analyzed in a reasonably general way using the techniques of effective field theory~\cite{Fichet2017, Brax2018}.

Neutron interferometry makes use of the neutron's quantum nature to probe both fundamental and nuclear physics. The electrical neutrality of the neutron coupled with its small magnetic moment, small neutron-electron scattering amplitude, and very small electric polarizability free it from the electromagnetic backgrounds faced by searches for exotic interactions which use test masses made of atoms. The related ability of slow neutrons to penetrate macroscopic amounts of matter and to interact in the medium with negligible decoherence also allows the quantum amplitudes governing their motion to accumulate large phase shifts that can be determined with high precision~\cite{Nico05b, Dubbers11, Pignol:2015}.

Perfect crystal neutron interferometry as technique is employed at one US (NIST) and one international neutron source (ILL).  This is mainly due to interferometry's requirement of having an environmental-decoupled space for long-term phase stability. The ILL's S18 beamline supports part-time currently the highest intensity neutron interferometer facility which has historically focused on studying aspects of quantum mechanics \cite{Sponar_2020,Denkmayr_2018,Sponar2021,Danner2020}.  At the NCNR, two full-time facilities are dedicated to perfect crystal neutron interferometry, one of which has the highest phase stability and fringe visibility in the world \cite{shahi_niof}.  

These features of slow neutron interactions have therefore been exploited in several searches for possible new weakly coupled interactions of various types, including chameleon dark energy fields, light $Z^{'}$ bosons, in-matter gravitational torsion and nonmetricity of spacetime, axion-like particles, and exotic parity-odd interactions~\cite{Leeb92, Bae07, Ser09, Ig09, Pie12, Yan13, Lehnert14, Lehnert15, Jen14, Lemmel2015, Li2016, Lehnert2017, Haddock2018b, Cronenberg2018}. A thorough review of almost all of the recent results from neutron measurements of this type has recently appeared~\cite{Sponar2021}, and another recent review~\cite{Safronova2018} has placed this work in the context of analogous investigations using atomic measurements.

The phase shift arising from dynamical diffraction in large perfect crystals now sets the best limits on Yukawa-modified short range gravity between 20~pm and 10~nm~\cite{Heacock21} and provides a determination of the neutron charge radius in a systematically independent way.  Understanding of the neutron charge radius has been recently improved both from effective field theory \cite{Filin20} and increasingly large nuclear data sets available \cite{Atac:2021b,Atac21}.  In addition, dynamical diffraction can be used to constrain exotic spin-dependent interactions~\cite{Horne88, Gentile19}.  Searches for spin and velocity-dependent interactions from spin-$1$ boson exchange~\cite{Piegsa2012} using slow neutron spin rotation~\cite{Haddock2018b} set the best limits for force ranges from mm to atomic scales. 

Neutron interferometry provides sub-percent scattering length data \cite{Schoen_2003,Huffman_2004,Haun_2020,Huber_2014} for low-Z isotopes for the bench marking of NN + 3NI nuclear models \cite{Pudliner_1997,Pieper_2001}.  In addition,  precise measurements of neutron scattering lengths helps constrain low energy constants used in building models at higher orders in chiral effective field theory \cite{Machleidt_2011}. 


Gravity resonance spectroscopy~\cite{Jenke2011, Abele2008}, which creates coherent superpositions of bound states of neutrons formed in a potential from the Earth's gravity and a flat mirror, which has been used to investigate several different types of exotic interactions~\cite{Ivanov2013, Jen14, Ivanov2016, Cronenberg2018, Klimchitskaya2019}, most recently CPT/Lorentz violation in the interactions of neutrons with the gravitational field of the Earth~\cite{Ivanov2021}. A qBOUNCE apparatus which implements vibrational Ramsey spectroscopy has seen its first signal~\cite{Sedmik2019}.   Instrumention appropriate for imaging these bound states is also under development at LANL UCN facility\cite{KUK2021}.

In the last few years, progress has been made in generating and detecting exotic quantum neutron states.   For instance, the recent demonstration at ISIS and ORNL of two- and three- variable single-neutron quantum entanglement in the spin, path, and energy qubits of polarized meV neutron beams~\cite{Shen2020, Kuhn2021} can open a new field of entangled neutron scattering. By manipulating the neutron phase one can generate orbital angular momentum (OAM) states that are of intellectual interest for possible applications in quantum sensing with neutrons and for neutron scattering studies of materials, where the nonzero $\vec{L}$ of the beam can selectively couple to certain topological excitations in condensed matter.   The methods of generating neutron OAM has evolved in just a few short years. Neutron OAM was first demonstrated using spiral phase plates inside a perfect crystal neutron interferometer at NIST~\cite{Clark2015}. Later using with orthogonal magnetic prisms~\cite{Sarenac2018, Sarenac2019} 
and more recently with fork-dislocation gratings \cite{Sarenac_2022}.  
Neutron OAM generation by strong nuclear and electromagnetic neutron spin-orbit interactions is also expected on theoretical grounds~\cite{Afanasev2019, Afanasev2021, Geerits2021} and is now under experimental investigation.  Both higher entangled states and OAM states provide new methodologies for nuclear study.

\section{Facilities for Fundamental Neutron Physics Research in the
  U.S.}

The fundamental neutron physics community faces special challenges. These diverse, often small scale, experiments have  varied needs which are not currently  met by any DOE Nuclear Physics funded host laboratory.  
In the U.S., the majority of fundamental neutron physics experiments have been performed at the Fundamental Neutron Physics Beamline (FNPB) at ORNL, the NIST Center for Neutron Research (NCNR), and the Ultracold Neutron Source at LANL. These experiments at non-DOE-NP operated facilities have provided a unique mutually-beneficial training ground for a entire generation of scientists in the field. As articulated above, the scientific impact enabled by these facilities/experiments is large, and much of the future work described requires continued and predictable access. Indeed, scientific progress is currently limited in part by the capability of these facilities to host experiments.
All three locations leverage resources funded by other agencies.   The construction and operation of the SNS, which provides cold neutrons to FNPB, is funded by the DOE BES.  The operation of the reactor and beamline development at the NCNR is funded by the Department of Commerce. The operation of the LANSCE accelerator, which provides proton beam to the LANL UCN source, is funded by the DOE NNSA. This, in combination with the implied reliance on research funds, necessarily presents risks to the fundamental neutron Nuclear Physics program. The three locations provide highly complementary capabilities, representing a high flux pulsed neutron source, a high flux cold beam, and high density UCN source, respectively. To fully realize the scientific program described above, improvements in the funding model should be made to ensure these US resources can continue to both support the full life-cycle of current and planned experiments and enhance their capabilities to support future improvements in precision.  
%

\subsection{Fundamental Neutron Physics Beamline at ORNL}

The FNPB is located at the SNS, whose 1.4\,MW accelerator provides the world's most intense source of pulsed neutrons. The SNS will be upgraded to 2\,MW power by the middle of the decade~\cite{Champion:2022inx} to support a planned Second Target Station. The SNS delivers 1.3\,GeV protons at 60\,Hz to a mercury target, releasing spallation neutrons which are subsequently moderated to cold neutrons, and serves 20 beamlines, including instruments primarily focused on material science studies and the FNPB. The FNPB on beamline 13 (BL13) is one of three instruments viewing the lower, downstream, 20\,K liquid hydrogen moderator. FNPB, which began operations very early 2010's, is the only beamline supported by DOE NP funded experiments, 
and is devoted to high impact studies of the fundamental properties of the neutron~\cite{Fomin:2014hja}. The primary beamline, BL13B, provides a polychromatic neutron beam that supports the main FNPB physics program, with typically one long-term experimental effort operating a time. The program commenced with precision studies of hadronic parity violation, including a first observation of the parity violating correlation between the neutron spin and $\gamma$ emitted after capture by the proton by the NPDGamma Collaboration~\cite{NPDGamma:2018vhh} and precision measurements of the parity violating asymmetry in neutron capture on $^3$He~\cite{n3He:2020zwd}, accomplished within the last LRP period. FNPB will primarily support a program of precision beta decay until late in the decade. Currently commissioning is the Nab experiment, which will measure unpolarized coefficients in neutron beta decay with unprecedented precision~\cite{Baes2014,Fry19}, proposed to be followed by pNab after minor modifications, to also access the polarized angular coefficients with excellent precision. The flagship of FNPB is the nEDM@SNS experiment to search for the neutron's EDM~\cite{Ahmed_2019}, which will begin commissioning and data-taking in the late 2020's, and is expected to dominate operations in the Long Range Plan period beyond this one. When inserted into the beamline, a double-crystal monochromator system provides 8.9\,\AA~neutrons to BL13A, which is used for early R\&D for nEDM@SNS in parallel with Nab data-taking, but cannot be operated during data-taking of the nEDM@SNS experiment (which requires the 8.9\,\AA~neutrons). The FNPB program includes very high priority experimental efforts for the field of fundamental symmetries, and increased operational and personnel support and protection of the research funding that provides it will be essential to meet the demands of the growing onsite efforts, ensure continuity and succession planning in critical areas of expertise, and to provide for workforce training and development, especially by capitalizing on efforts to identify talent from a more diverse community.

\subsection{NIST Center for Neutron Research}

Within the US Department of Commerce, the National Institute of Standards and Technology (NIST) operates the NIST Center for Neutron Research (NCNR).  The NCNR's mission is to provide neutron measurement capabilities to the broad U.S. research community.
Based around a 20 MW research reactor it is a national center for research using thermal and cold neutrons, offering its instrumentation for use by both industry and academia. Many of these instruments rely on intense beams of cold neutrons emanating from an advanced liquid hydrogen moderator.
Since the early 1990s the NIST Physical Measurement Laboratory and the NCNR together have operated two high-flux cold neutron beams and multiple monochromatic cold neutron beams that have historically been used for fundamental science. A curved ballistic beamline designed by NIST specifically to support future precision measurements, NG-C, and it's predecessor NG-6 have been used for a series of precision beta decay experiments, measurements of hadronic parity violation, and searches for T-violation, see for example~\cite{Yue:2013a,PhysRevC.83.022501, PhysRevLett.107.102301}.   Two of the monochromatic beamlines support interferometry facilities that provide the highest fringe visibility and phase stability in the world.  These interferometer facilities have been used to measure nuclear scattering lengths of light elements and set limits on BSM physics, see for example~\cite{doi:10.1126/science.abc2794}.  Within the period of the last LRP significant advances were made using NIST beamlines. These range from a first precision measurement of neutron radiative decay and new asymmetry measurements to scattering length measurements and new interferometry techniques~\cite{PhysRevC.103.045502,HEACOCK2020893,PhysRevLett.124.012501,PhysRevC.100.015204,PhysRevC.100.034005,PhysRevLett.121.183602,Yue2018PrecisionDO,PhysRevLett.120.113201,Darius2017MeasurementOT,Bales2016PrecisionMO}.  Currently a second run of the beam-based neutron lifetime experiment that will weigh in on the current neutron lifetime discrepancy and new measurements of scattering lengths are in progress.

As detailed in~\cite{coldsourceupgrade}, sometime in the 2024 time frame the
current liquid H$_2$ cold source will be replaced with a liquid D$_2$ cold source. This will result in a factor of two increase in usable flux on NG-C, making it equivalent to the highest flux neutron beams in the world, the only cold beam of close to this intensity available in North America, and the only one operated such that it can support the extended run times necessary for high-statistics precision beta decay experiments. A third interferometer facility utilizing phase gratings with meter-long path length will also be installed. Despite being operated by the Department of Commerce, NIST has supported a number of DOE and NSF funded experiments, providing significant research scientist and engineering support, as well as materials and supplies. Direct investments by DOE are highly leveraged and explicitly recognized as such.  Despite previous success, the current situation provides significant and important challenges. Shielding and engineering costs are significantly higher with the primary factor being safety concerns as the apparatuses become more complicated and as they operate in much higher radiation fields, this affects 
the ability to support projects throughout their lifecyle. Further investment is needed to ensure the continued ability of the community to fully utilize this important resource.

\subsection{Ultracold Neutron Source at LANL}
The LANL Ultracold Neutron (UCN) Source is based on a solid deuterium (SD$_2$) converter driven by spallation neutrons~\cite{Saunders:2013RSI}. The SD$_2$ converter produces UCN from cold neutrons via the so-called superthermal process. The technique was first demonstrated at LANL~\cite{MOrris2002}. This pioneering work led to the realization of the PSI UCN source~\cite{Anghel2009}, the TUM source~\cite{FRMII}, and a full-scale source subsequently built for the UCNA experiment (see e.g. Ref.~\cite{Brown2018}). 
The LANL UCN source went through a major upgrade funded by LANL LDRD funds in 2014-2018~\cite{Ito2018}. So far, it has provided UCN to the UCNA, the UCN$\tau$ (see e.g. Ref.~\cite{Gonzalez2021}), R\& D for the nEDM@SNS experiment, and searches for exotic neutron decays~\cite{Tang2018,Sun2018}. Currently, the UCN$\tau$+ experiment (an upgrade to the UCN$\tau$ experiment), the UCNA+ experiment (an upgrade to the UCNA experiment), the LANL nEDM experiment~\cite{Ito2018}, and the UCNPro$\beta$e experiment are being developed to run using UCN from the LANL UCN source. Development of various UCN detectors for current experiments and possible future experiments, including 2D imaging detectors~\cite{Wei:2016a,KUK2021}, is also performed.
Techniques to perform UCN interferometry for femto-eV energy resolution are being developed~\cite{Wang2020, KUK2021}.
Further improvement of the UCN source performance is possible by placing low enriched uranium between the spallation target and the cold moderator~\cite{Pattie2020}.

As of this writing, the LANL source is the only UCN source in North America that provides UCN to physics experiments. It is one of the world's best performing UCN source, hosting world leading UCN based experiments.  Its operation, however, is not supported by the DOE NP as a user facility. The cost of its operation has been covered by the funds to operate each experiment that is supported by the DOE NP (currently the UCN$\tau$ experiment), along with the LANL LDRD program that provides seed funding to support development of new experiments. As a result, despite a large number of requests
(both from inside and outside the U.S.) to use UCN from the LANL UCN source for various experimental efforts,
there is no mechanism to accommodate such requests, let alone supporting a user program.
Under the current funding model, it has been difficult to operate the source in a stable manner to support all the approved experiments, execute adequate maintenance, and perform developments for necessary improvements. To maintain leadership, the U.S. fundamental neutron physics community needs a UCN source that is adequately funded for stable operation and steady improvements.

It should also be mentioned that a new UCN source based on a SD$_2$ converter coupled to a reactor neutron source is being constructed at the PULSTAR reactor at North Carolina State University~\cite{Korobkina2014}. The primary purpose of this UCN source is to host an apparatus to study systematic effects for the nEDM@SNS experiment.

\subsection{Next generation UCN sources in the U.S.}
Neutrons represent an excellent opportunity to push the discovery potential of the precision frontier, with experimental efforts being primarily limited by number of available neutrons in terms of statistical sensitivity and ability to characterize systematic effects. Therefore, the US community's ability to remain competitive and enhance its science reach beyond this Long Range Plan depends crucially on research and development for brighter sources of neutrons. UCN have become an important emerging tool in neutron research, as evidenced by the number of UCN sources operational or being developed worldwide (see Tab.~\ref{tab:UCNsources}). This competitive environment provided strong motivation for the source upgrade implemented at LANL in 2014-2018, where a factor of five increment in UCN density was achieved, setting the stage for the world-leading UCNtau experiment and the LANL EDM experiment.  Over the past decade, however, the PSI SD$_2$ source and ILL liquid He source have also been steadily improving and new sources are approaching operational status.  To maintain our leadership in this field, continued investment in UCN source development is essential, which will to ensure a vibrant community that can continue to carry out high-impact science in fundamental neutron physics for the coming decade and beyond. The US UCN community has already started to work on ideas for the next-generation UCN sources. Below are some ideas that are being considered, which ranges from those that are close to being ready to be implemented to those that require significant R\&D and design work:

\begin{itemize}
\item {\bf Uranium neutron multiplier for the LANL UCN source:} Low-enriched uranium placed between the spallation target and the cold moderator can increase the UCN source output. The low-enriched uranium multiplies the fast neutrons from the spallation target. Preliminary studies show that an increase of UCN output up to 10 times may be possible~\cite{Pattie2020}. 

\item {\bf LHe converter based UCN source coupled to a spallation target:} By coupling a high-current spallation target to subcooled helium, it is expected to significantly enhance the UCN yields beyond the capabilities of current UCN sources and increase the density of UCN by several orders of magnitude to reach several thousands per cubic centimeters~\cite{Leung2019}. LANL Area A is being considered to be a possible site to house such a source, and a third target station at the SNS is another possibility.

\item {\bf LHe converter based UCN source at HFIR at ORNL:} The planned upgrade of the HFIR pressure vessel offers a unique opportunity to include design requirements for a UCN source based on a LHe converter coupled to cold neutrons. A preliminary study concluded that gains in UCN density of 2-3 orders of magnitude is possible with a UCN converter inhabiting one of the main beam ports as close as possible to the reactor, if the new pressure vessel design accommodated a larger diameter beam tube. Further studies are needed to optimize heat loads and minimize impact to isotope production at HFIR.

\item {\bf Future UCN source at NIST:} NIST is in the very early stages of designing a next-generation reactor to replace the existing NBSR.  While no concrete plans currently exist, this represents a once-in-a-generation opportunity to optimize a UCN source from the early design stages of this future facility. Unexplored options for utilizing the existing thermal column or a cold neutron beam based UCN source also exist at NIST in the nearer term.  
\end{itemize}

The UCN community in the U.S. will organize a series of workshops to discuss the optimum technology, compare various options including those mentioned above, select possible sites for the next generation UCN source in the U.S., and will perform necessary development of the next generation UCN source for the community as a whole. Modest support in the next decade for evaluation of these possibilities and design is critical to avoid missing a key opportunity for a world-leading next-generation UCN source in the US.

\begin{table}
    \centering
    \begin{tabular}{l|l|l|l|l|l} \hline\hline
    UCN source & Location & Converter & Neutron source & Status & Ref.\\ \hline
    ILL (Turbine) & ILL & Receding blades & Reactor & Operational & \cite{Steyerl1986} \\
    PSI & PSI & SD$_2$ & Spallation & Operational & \cite{Anghel2009,Becker2015,Bison2020}\\
    Mainz & Mainz & SD$_2$ & Reactor & Operational & \cite{Lauer2013}\\
    J-PARC & J-PARC & Decellerating mirrors & Spallation & Operational & \cite{Imajo2016}\\
    LANL & LANL & SD$_2$ & Spallation & Operational & \cite{Ito2018}\\    \hline
    SuperSUN & ILL & LHe & Reactor & Commissioning & \cite{Zimmer2016} \\
    PULSTAR & NCSU & SD$_2$ & Reactor & Commissioning & \cite{Korobkina2014} \\ \hline
    TRIUMF & TRIUMF & LHe & Spallation & Under development & \cite{SAhmed2019} \\
    TUM & TUM & SD$_2$ & Reactor & Under development & \cite{FRMII} \\
    PIK & PIK & LHe & Reactor & Under development & 
    \\ \hline \hline
    \end{tabular}
    \caption{UCN sources that are either operational or under development}
    \label{tab:UCNsources}
\end{table}

\section{Workforce development}


Since the last LRP, 17 new faculty lines in the US academic institutions and 10 new research scientists in the US National Laboratories have been created, signifying the high-profile, high-impact science of the fundamental neutron physics and the resulting growth of the community. There have also been faculty research groups ($\sim 5$ outside the FNP community) expanding their research portfolios to join collaborative work with neutrons. The diversity in project scale in the field of neutron physics is an important aspect for workforce development.  Early career scientists can develop leadership and project management skills on shorter term, smaller scale experiments with quicker turnaround to science results and opportunities for community visibility. Meanwhile, participation on larger scale, longer term experiments ensures a stable research program and eventual higher impact recognition. 
To support the challenges of the diverse, high-impact research portfolio presented here over the coming decade, it is of paramount importance to prepare next-generation researchers with specialized skills and intellectual capacities. 
The community has been organizing Fundamental Neutron Physics Summer Schools (every 3 years, though COVID pandemic interrupted the last planned school) to provide training of PhD students and postdocs. To sustain the anticipated growth, the community needs to recruit proactively from broader populations to include non-traditional and under represented populations.

\section{Diversity, equity, and inclusion}

The Fundamental Neutron Physics community shares the value that diversity and inclusion are essential for the future of our field, physics, science, and society. A diverse community that covers the arc of human potential from high school students to early career scientists brings talent, ideas and perspective that are a benefit to our research goals and a path to equity. The DOE 2020 Nuclear Physics Workforce Survey reports that since 2004, about 25\,\% of PhDs granted in nuclear physics were to members of underrepresented groups, but fewer than 2\,\% to Black or African Americans and 4\,\% to Hispanic or Latino nuclear physicists. These survey results are consistent with physics as a whole, but far from representative of the talent pools in the US. 

The Fundamental Neutron Physics community has much to offer to help lead the way to a more diverse Physics work force: Our research is undertaken by relatively small teams (a dozen researchers is typical) and is characteristically multidisciplinary, drawing on diverse talents. Our projects provide the opportunity for mentoring students more personally with more frequent contact, and the training provides a broad range of employment opportunities beyond staying in the field. Our projects also have many opportunities for recruitment of undergrad and even high-school students who can contribute. In addition to incentives from university PIs, all three labs have a number of programs that can support high school and undergrad students (SHIP and SURF at NIST, SULI at DOE National Laboratories/Facilities including ORNL and LANL for undergrad students and postbach students, high school internship program at LANL). 
The University of Tennessee, Knoxville and ORNL have partnered to develop a new year-long student internship program, Nuclear Physics in Eastern Tennessee Fellowships, which provides professional development and research experience in nuclear physics for a cohort of undergraduate students from Minority Serving Institutions. 

\section{Priorities and Opportunities for Fundamental Neutron Physics}

The Fundamental Neutron Physics community has made significant strides since the last LRP, with world-leading results in neutron physics including measurements of the neutron lifetime, beta-decay correlations and hadronic parity violation as well as the development of new high-impact experiments including BL3, Nab, LANL nEDM and nEDM@SNS. In order to maintain this exciting program and capitalize on past investments, the community has identified a number of priorities and opportunities for the field covering the time-frame of the LRP under development. These are:

\vskip .1in

$\hskip .1in\bullet$ Funding for completion of the construction of the nEDM@SNS apparatus and the start of data-taking

\vskip .05in

The nEDM@SNS has achieved a number of significant technical milestones and is constructing and commissioning major components for the experiment. Continued support to complete these components is needed. In addition, several major capital acquisitions are required including an extension of the cold-neutron guide and the construction of a new end-station to house the experiment at the end of the FNPB beamline. Completion of the apparatus and subsequent data acquisition and analysis will herald the start of a next-generation of nEDM searches.

\vskip .1in

$\hskip .1in\bullet$ Investment in additional new funding to support research and beamline operations, including additional personnel, for FNPB, the NIST fundamental neutron physics beamlines, and the LANL UCN source to run experiments, improve capabilities and provide continuity

\vskip .05in

These three complementary capabilities provide the most intense pulsed beam, cold beam and UCN source in the US. Together they host all the primary experiments to improve the precision of searches for CKM non-unitarity and BSM physics. Such an investment is required to enhance the productivity of these experiments by providing focused support to help with operation of the experiments. It is also needed to provide flexibility for exploring new initiatives, help mentor and train the onsite early-career workforce for the next generation experiments, enhance safety and compliance and provide continutity for these generally long-duration experiments. 

\vskip .1in 

$\hskip .1in\bullet$ Increased support for theoretical groups that are involved in all aspects of fundamental neutron physics research, from phenomenology, to effective field theories, hadronic physics and lattice QCD, expanding connections with the high-energy physics and nuclear structure communities

\vskip .05in

The interpretation of low-energy  searches of BSM physics requires a multi-scale and multi-disciplinary approach that integrates expertise from several branches of nuclear and high-energy theoretical physics: from model building and phenomenology, to lattice QCD, to hadronic EFTs for few- and many-body systems.
The recent progress on nEDM, PV couplings  and $\gamma W$-box calculations from lattice QCD and EFT for radiative corrections to pion, kaon and neutron decays are very promising, but the small number of groups involved highlights the need to grow this community. The challenges posed by the two-nucleon calculations needed to understand hadronic CP- and P-violation require a new level of effort. There is urgent need to expand the EFT and lattice QCD communities that work on Fundamental Symmetries, and to leverage the connections with nuclear structure and high-energy physics. Possible options are new topical collaborations/theory initiatives focused on Fundamental Symmetries, in particular to provide support for students, postdocs and junior faculty to grow the theoretical workforce.

\vskip .1in 





$\hskip .1in\bullet$ Develop new funding mechanisms to support R\&D activities focused on future experiments and capabilities, for example a next-generation UCN source and high flux, high polarization and high uniformity neutron beam polarizers for cold neutron beams

\vskip .05in

In the absence of a mission-centered facility focused on operating fundamental neutron physics experiments, developing future experiments and capabilities can be challenging. Future increases in sensitivity necessitate new purpose-specific directions in detector technology, improved metrological tools specifically for neutrons, and improved capabilities to both produce and manipulate neutrons.   For comparison, the path to the Electron-Ion Collider was greatly facilitated by focused effort from both JLab and RHIC.

\bibliography{neutron_whitepaper_refs}
  
\end{document}